\documentclass[
 reprint,
 superscriptaddress,
 amsmath,amssymb,
 aps,
]{revtex4-2}

\usepackage{graphicx}
\usepackage{dcolumn}
\usepackage{bm}
\usepackage[colorlinks=true, citecolor=Blue, linkcolor=Navy, urlcolor=RoyalBlue]{hyperref}
\usepackage[mathlines]{lineno}

\usepackage{multirow} 
\newcolumntype{P}[1]{>{\centering\arraybackslash}p{#1}}
\newcolumntype{M}[1]{>{\centering\arraybackslash}m{#1}}

\usepackage{diagbox}
\usepackage[svgnames,table]{xcolor}

\usepackage{caption}
\usepackage{subcaption}
\usepackage{ragged2e}

\usepackage{braket}

\DeclareCaptionJustification{justified}{\justifying}

\newcommand{\HG}{\mathrm{HG}}
\newcommand{\HGmode}[2]{\ensuremath{\HG_{#1,#2}}}

\newcommand{\LG}{\mathrm{LG}}
\newcommand{\LGmode}[2]{\ensuremath{\LG_{#1,#2}}}

\newcommand{\HGzero}{\HGmode{0}{0}}
\newcommand{\HGthree}{\HGmode{3}{3}}
\newcommand{\LGtwo}{\LGmode{2}{2}}

\begin{document}


\title{Thermal Deformation Reduction in High-Power Interferometry with Higher-Order Laser Modes}

\author{Liu~Tao}
\email{liu.tao@apc.in2p3.fr}
\affiliation{Universit\'e Paris Cit\'e, CNRS, Astroparticule et Cosmologie, F-75013 Paris, France}
\affiliation{University of Florida, 2001 Museum Road, Gainesville, Florida 32611, USA}

\author{Yuhang~Zhao}
\affiliation{Institute for Gravitational Wave Astronomy, Henan Academy of Sciences, Zhengzhou, 450046, China}

\author{Zong-Hong~Zhu}
\affiliation{School of Physics and Astronomy, Beijing Normal University, Beijing 100875, China}

\affiliation{Institute for Frontiers in Astronomy and Astrophysics, Beijing Normal University, Beijing 102206, China}

\author{Paul Fulda}
\affiliation{University of Florida, 2001 Museum Road, Gainesville, Florida 32611, USA}

\date{\today}

\begin{abstract}
Test-mass thermal noise is a limiting noise source for current and next-generation ground-based gravitational-wave observatories in their most sensitive frequency band. In addition to ongoing efforts to reduce coating mechanical loss, complementary approaches such as the use of uniform-intensity laser beams, including higher-order Laguerre-Gaussian (LG) and Hermite-Gaussian (HG) modes, have been proposed as alternatives to the fundamental Gaussian beam due to their thermal-noise advantages. As the interferometer operating power increases toward the megawatt regime, as anticipated for next-generation detectors, thermal aberrations arising from absorption in the high-reflectivity coatings of the test masses become an increasingly severe challenge. In parallel, the spatially more extended intensity profiles of higher-order modes, which enable their thermal-noise benefits, modify the thermal loading of the test masses, making it essential to assess their behavior under absorption-induced self-heating. In this work, we quantify the robustness of higher-order incident beams against thermal deformation. We show that, under identical operating conditions, higher-order modes produce significantly more uniform thermally induced mirror distortions compared to the sharply-peaked aberrations generated by the fundamental mode. As a result, substantially less thermal compensation power is required for their active correction, with the required optimal curvature correction reduced to 33\% of that of the fundamental mode for \LGtwo{} mode, and to 24\% for \HGthree{} mode. In addition, we show that the residual thermal deformation for higher-order modes leads to significantly lower optical power loss and enables larger power buildup and enhanced modal purity in an aLIGO-like optical cavity. We further demonstrate that astigmatism compensation enhances the intracavity modal purity of HG modes under self-heating-induced thermal deformation. These results highlight that higher-order laser modes mitigate thermal noise while intrinsically suppressing beam self-heating-induced thermal distortions, making them more thermally robust and well-suited for operation in high-power gravitational-wave interferometers.

\end{abstract}

\maketitle


\section{Introduction\label{sec-intro}}
\begin{figure*}[t]
    \centering
        \includegraphics[width=1\linewidth]{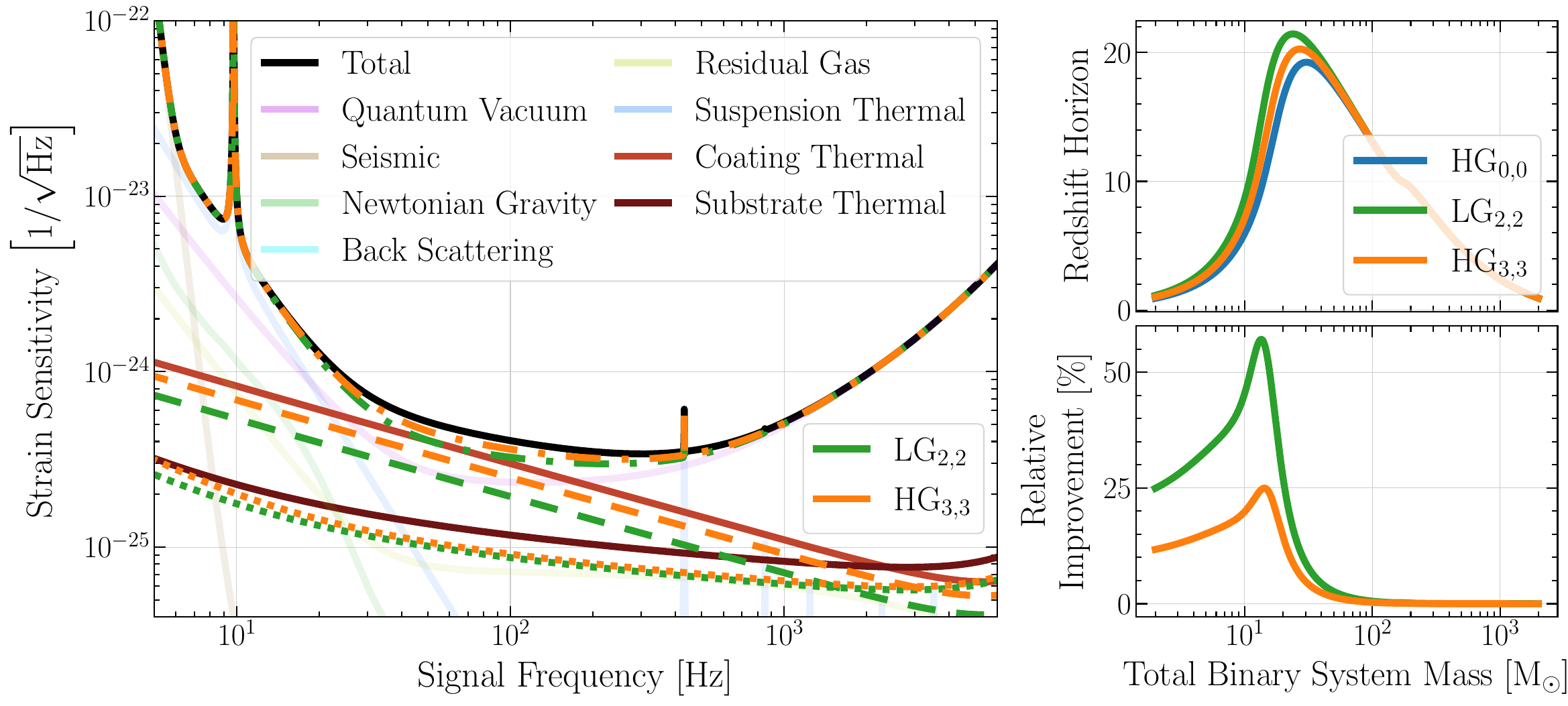}
    \caption{Noise budget for the Einstein Telescope high-frequency interferometer using incident beams in the fundamental \HGzero{} mode and the higher-order \LGtwo{} and \HGthree{} modes. \textit{Left:} Strain sensitivity curves, with the total sensitivity shown in black and individual noise contributions indicated in color. In the most sensitive band around 100~Hz, the sensitivity is dominated by test-mass coating thermal noise (red). The thermal noise for the \LGtwo{} (green) and \HGthree{} (orange) modes is also shown, with dashed and dotted lines indicating the coating and substrate contributions, respectively, while the total sensitivities for the higher-order modes are shown as dot-dashed lines. \textit{Right:} Cosmological redshift detection horizons for nonspinning, equal-mass binary black hole mergers as a function of the total source-frame mass (top), together with the relative improvement for using the \LGtwo{} and \HGthree{} modes (bottom).}
    \label{fig-sensitivity}
\end{figure*}

Thermal noise of the test masses, arising from the Brownian motion of molecules in both the bulk substrate and the reflective coatings, poses a fundamental limitation for current gravitational-wave (GW) observatories such as Advanced LIGO (aLIGO) and Virgo around 100~Hz, the frequency band where they achieve the maximum sensitivity to astrophysical sources such as binary black hole mergers~\cite{Capote_2025, Acernese_2015}. In addition, for future third-generation detectors, including Cosmic Explorer and the Einstein Telescope, thermal noise is expected to remain the dominant noise source in the most sensitive frequency band~\cite{ETDesignReportUpdate2020, CEHorizonStudy}. For instance, as illustrated in Fig.~\ref{fig-sensitivity} for the Einstein Telescope high-frequency interferometer, the total strain sensitivity (black) is dominated by coating thermal noise (red) from roughly 30 to 200~Hz. In addition, thermal noise remains one of the dominant noise sources limiting the stability and accuracy of laser-based optical experiments that rely on stabilized cavities~\cite{PhysRevLett.93.250602}, including applications in optical atomic clocks~\cite{Oelker:2019kqe}, atom interferometry~\cite{DovaleAlvarez:2019ugw}, and searches for dark matter~\cite{Savalle_2021}.

Significant effort has been devoted to mitigating thermal noise through improvements in detector design, including the development of coating materials with lower mechanical loss and the implementation of cryogenic operation with silicon test masses using longer-wavelength lasers~\cite{PhysRevLett.127.071101, Adhikari_2020}. Complementary to these approaches, an alternative strategy is to employ laser beams with more uniform intensity profiles, in place of the currently used fundamental Gaussian beam. According to the fluctuation-dissipation theorem, such beam profiles can more effectively average thermally induced fluctuations across the test mass surface and substrate, potentially reducing the impact of thermal noise without requiring modifications to the test masses themselves~\cite{Mours_2006, PhysRevD.82.042003}.

\begin{figure}[b]
    \centering
    \includegraphics[width=1\linewidth]{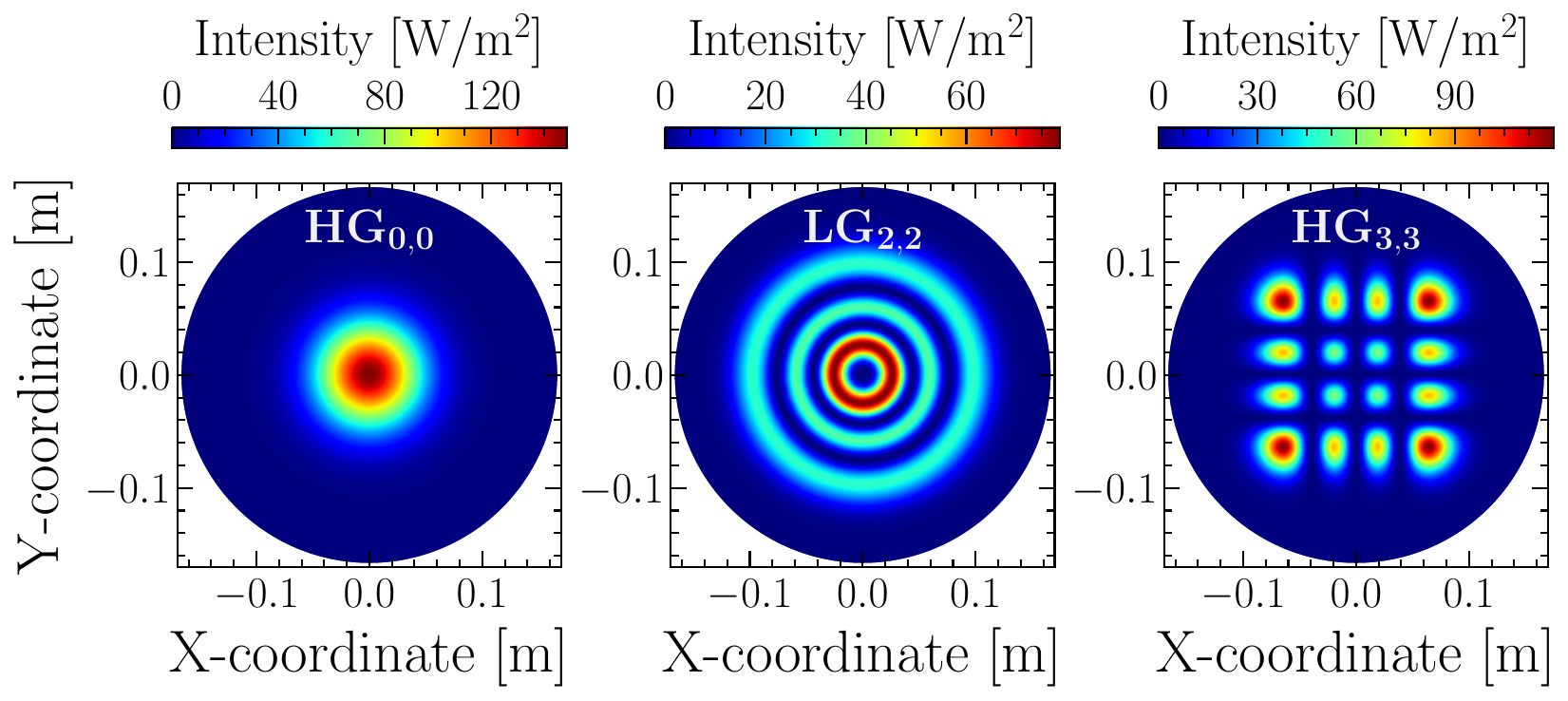}
    \caption{Intensity profiles of the fundamental \HGzero{} mode, and the sixth-order \LGtwo{} and \HGthree{} modes. All beams are normalized to 1~W, with beam sizes rescaled to yield 1~ppm clipping loss on an aLIGO-like test mass.}
    \label{fig-intensity_homs}
\end{figure}

A widely studied family of such uniform-intensity laser beams includes higher-order Laguerre-Gaussian (LG) and Hermite-Gaussian (HG) modes. Fig.~\ref{fig-intensity_homs} captures the intensity distributions of the unit-Watt \LGtwo{} and \HGthree{} modes, both of sixth order, compared with the fundamental Gaussian mode. Fig.~\ref{fig-sensitivity} compares the total strain sensitivity of the Einstein Telescope high-frequency interferometer if the currently used fundamental Gaussian beam is replaced by the higher-order \LGtwo{} (green) and \HGthree{} (orange) modes; the corresponding total sensitivities are shown as dot-dashed lines. All technical noise sources except coating and substrate thermal noise are displayed in the background with reduced opacity to better highlight the thermal noise. The reductions in coating and substrate thermal noise associated with higher-order modes assume the same finite-sized mirrors, with the beam sizes rescaled to maintain the same clipping loss, and the respective contributions are shown as dashed and dotted lines, respectively~\cite{PhysRevD.82.042003}. The right panels of Fig.~\ref{fig-sensitivity} present the resulting cosmological redshift detection horizons for nonspinning, equal-mass binary black hole mergers as a function of the total source-frame mass for the fundamental and the two higher-order modes (top), together with the relative improvement with respect to the fundamental Gaussian mode (bottom). As indicated, the use of \LGtwo{} and \HGthree{} modes can yield up to 57\% and 25\% improvements, respectively, in the sky-averaged detection range for binary mergers due to the associated reduction in thermal noise.

The potential implementation of the higher-order LG and HG modes in interferometric GW detectors has been studied extensively through both numerical and tabletop experimental demonstrations~\cite{PhysRevD.82.012002, bond2011, Sorazu_2013, PhysRevLett.105.231102, PhysRevD.90.122011, PhysRevD.92.102002, Tao2020, PhysRevD.103.042008, PhysRevLett.132.101402}. For instance, despite their vast thermal noise advantages, LG modes have been shown to exhibit significant coupling to degenerate modes of the same order in the presence of asymmetric distortions with azimuthal dependence~\cite{bond2011}. On the other hand, despite their more modest thermal noise benefits compared to LG modes, HG modes have been demonstrated that they can be robust enough against mirror aberrations by intentionally breaking the pseudo-degeneracy of HG modes of the same order with astigmatism~\cite{Tao2020}. In addition to their enhanced robustness against mirror scattering errors, higher-order HG modes have been shown experimentally to be generated with high efficiency at high power levels~\cite{10.1063/5.0137085}. They are also compatible with squeezed-light injection techniques for quantum-noise reduction of up to 10~dB, demonstrating that their thermal-noise advantages can be realized without compromising quantum-noise performance~\cite{PhysRevLett.128.083606, PhysRevLett.129.031101}.

Beyond their thermal noise advantages in laser interferometers, higher-order LG and HG modes are an active area of research across a variety of applications. These include increased data capacity in free-space optical communications~\cite{HOM_nature}, high-resolution manipulation and characterization of topologically complex electronic materials~\cite{Lee2019}, precision imaging and object identification for remote sensing~\cite{PhysRevLett.110.043601}, enhanced small-displacement measurements in precision metrology~\cite{10.1063/1.4869819, h91l-w8w7}, and high-efficiency multimode quantum information processing for quantum optical communications~\cite{PhysRevLett.98.083602}.

While higher-order modes offer clear advantages in mitigating thermal noise, operating interferometers at the elevated circulating powers expected for next-generation detectors, in excess of one megawatt, introduces additional challenges associated with thermal absorption and induced wavefront aberration~\cite{PostO5Report:2022, LIGOwhitepaper2025}. Even small amounts of laser power absorbed in the high-reflectivity (HR) coatings of the test masses produce significant temperature gradients across the substrate, leading to location-dependent changes in the optical path length~\cite{Wang_2017}. This thermorefractive substrate lensing can modify the input beam parameters, causing imperfect mode matching and associated optical losses. In addition, thermal stress resulting from absorption can induce thermoelastic deformation of the test mass surfaces, directly altering the cavity eigenmodes and further degrading interferometer performance.

To preserve optimal coupling between the laser and the interferometer, thermal compensation systems are employed to correct these thermal aberrations and restore the test mass to its nominal optical state~\cite{Brooks_2016}. However, as operating powers approach the megawatt regime, both the required compensation power and system complexity increase substantially~\cite{Rosauer:25}. In this work, we demonstrate that, for the same reason that gives rise to their thermal noise advantage, higher-order laser beams, including both LG and HG modes, distribute thermal loads more evenly across the test-mass surface due to their broader intensity profiles, resulting in significantly reduced thermal aberrations compared to the fundamental Gaussian beam~\cite{Vinet_2007}. Consequently, they require substantially less actuation power and complexity in the thermal compensation and can maintain near-optimal intracavity power buildup and modal purity, making them significantly more thermally robust and suitable for next-generation high-power GW detectors.

The paper is organized as follows. In \S\ref{sec-background}, we summarize the steady-state thermal response of test masses due to absorption in the high-reflectivity coating. In \S\ref{sec-thermal}, we present finite element analysis results and analyze the thermally induced wavefront aberration maps for both the reflected and transmitted fields of the test mass. In \S\ref{sec-optical}, we examine the optical performance of an aLIGO-like symmetric cavity for the fundamental \HGzero{} and higher-order \LGtwo{} and \HGthree{} modes, considering single-bounce optical loss, required optimal curvature correction, cavity power buildup, and modal purity. Finally, \S\ref{sec-conclusion} concludes with an overview of the broader implications of this work and potential directions for future research.

\section{Background \label{sec-background}}
Here, we briefly review the thermal response of test masses subjected to absorption at the HR coating~\cite{Vinet2009zz}. Since we consider absorption from arbitrary laser beams, including higher-order Hermite-Gaussian modes that generally lack cylindrical symmetry, we adopt Cartesian coordinates $(x,y,z)$ to describe the thermal response. We consider a cylindrical optic of radius $R$ and thickness $H$, with $x$ and $y$ denoting the transverse coordinates of the incident beam and $z$ representing the longitudinal propagation direction along the test mass. The coordinate ranges are defined as $0 \leq x,y \leq R$ with $x^2+y^2 \leq R^2$ for the transverse plane, and $-H/2 \leq z \leq H/2$ longitudinally. Thermal absorption occurs at the HR surface, located at $z = H/2$.

At thermal steady state, as the cavity is held on resonance, the time-dependent heat transfer equation reduces to the Laplace equation:
\begin{equation}
\nabla^{2} T\left(x, y, z\right)=0 ,
\label{equ-Laplace}
\end{equation}
where $T\left(x, y, z\right)$ is the temperature rise from the ambient temperature $T_{0}$. In general, $T$ depends on $(x, y, z)$ since we consider incident heat flux in the form of arbitrary higher-order mode beams that in general do not have cylindrical symmetry.

The boundary conditions describe the heat flows on the surfaces and edges of the test mass. At thermal equilibrium, we have the following balance on the irradiated surface
\begin{equation}
\begin{aligned}
&-K \left[\frac{\partial T(x, y, z)}{\partial z}\right]_{z=H / 2} \\&
=\epsilon_{\mathrm{f}} \sigma_{\mathrm{SB}}\left(\left[T_{0}+T(x, y, H/2)\right]^{4}-T_{0}^{4}\right) - I_{\mathrm{beam}}(x, y) ,
\end{aligned}
\label{equ-BC1}
\end{equation}
where $K$ denotes the thermal conductivity and $\sigma_{\mathrm{SB}}$ is the Stefan-Boltzmann constant ($\sigma_{\mathrm{SB}} \sim 5.670 \times 10^{-8}\,\mathrm{W\,m^{-2}\,K^{-4}}$). $\epsilon_{\mathrm{f}}$ is the emissivity of fused silica at the front surface. The first term on the right-hand side represents the power flow of thermal radiation according to the Stefan-Boltzmann law, while the second term, $I_{\mathrm{beam}}(x, y)$, corresponds to the intensity of the laser beam incident on the HR surface. We consider beams in the form of \HGzero{}, \LGtwo{}, and \HGthree{}, with $I_{\mathrm{beam}}(x, y)$ given by the intensity distributions shown in Fig.~\ref{fig-intensity_homs}.

\begin{figure*}[t]
    \centering
    \begin{subfigure}{0.49\textwidth}
        \centering
        \includegraphics[width=1\linewidth, trim={30mm 10mm 15mm 20mm}, clip]{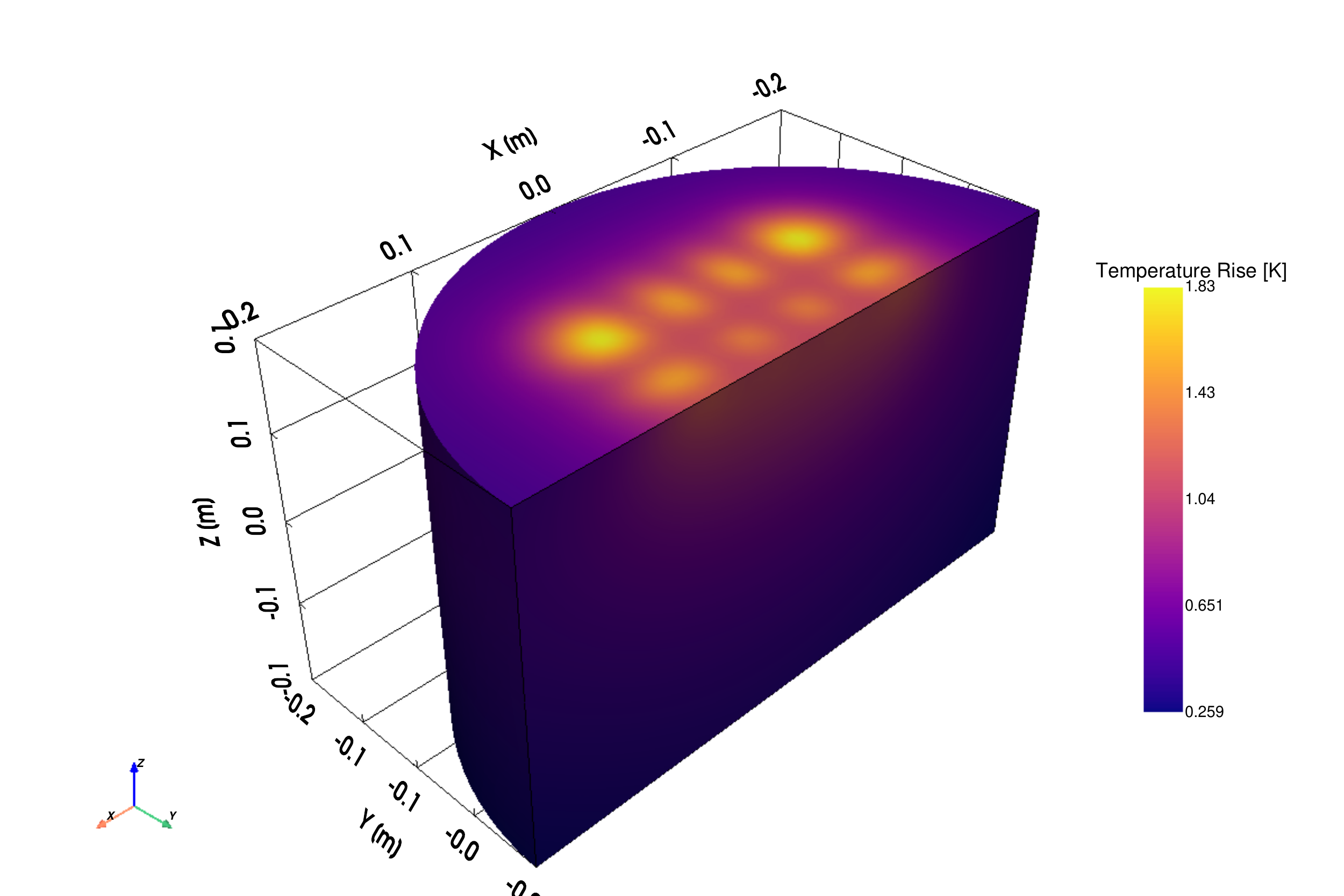}
        \caption{Temperature rise}
    \end{subfigure}
    \hfill
    \begin{subfigure}{0.49\textwidth}
        \centering
        \includegraphics[width=1\linewidth, trim={30mm 5mm 20mm 35mm}, clip]{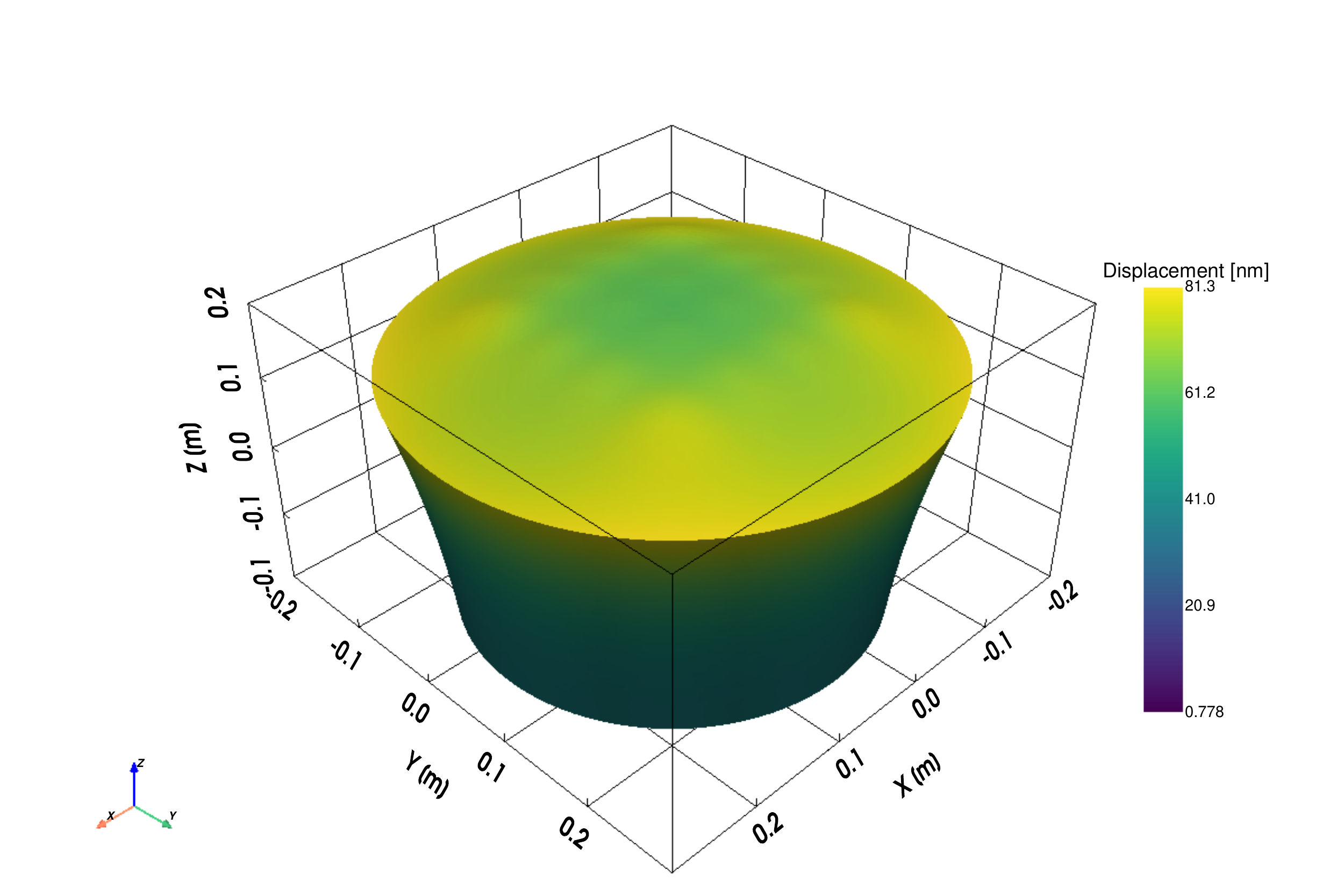}
        \caption{Thermoelastic deformation}
    \end{subfigure}

    \caption{Thermal finite element analysis performed with \texttt{FEniCSx} for an aLIGO-like test mass subjected to coating absorption from a unit-Watt \HGthree{} beam incident on the front surface ($z = 0.1$~m). \textit{Left:} temperature rise gradient within the substrate; \textit{Right:} resulting thermoelastic deformation of the test mass.}
    \label{fig-FEA_3D}
\end{figure*}

Assuming a small temperature rise, $T(x, y, z) \ll T_{0}$, we can further linearize the right hand side of Eq.~(\ref{equ-BC1})
\begin{equation}
\begin{aligned}
&-K\left[\frac{\partial T(x, y, z)}{\partial z}\right]_{z=H / 2} \\&
=4 \epsilon_{\mathrm{f}} \sigma_{\mathrm{SB}} T_{0}^{3} T(x, y,H / 2) - I_{\mathrm{beam}}(x, y) .
\end{aligned}
\label{equ-BC2}
\end{equation}

Similarly, on the opposite surface at $z = -H/2$, the radiation flow is reversed, giving
\begin{equation}
K\left[\frac{\partial T(x, y, z)}{\partial z}\right]_{z=-H / 2}=4 \epsilon_{\mathrm{f}} \sigma_{\mathrm{SB}} T_{0}^{3} T(x, y, -H / 2) ,
\label{equ-BC3}
\end{equation}
whereas along the barrel on the cylindrical test mass, we have
\begin{equation}
\begin{aligned}
- K \frac{\partial T(x,y,z)}{\partial n}
&= 4 \epsilon_{\mathrm{edge}} \sigma_{\mathrm{SB}} T_0^{3}\, T(x,y,z),
\\[0.3em]
\text{on}\quad & x^2 + y^2 = R^2 .
\end{aligned}
\label{equ-BC4}
\end{equation}
Here, $\hat{n}$ denotes the radial normal of the test mass, and we allow the surface emissivity of the barrel, $\epsilon_{\mathrm{edge}}$, to differ from that of the front surface, $\epsilon_{\mathrm{f}}$, to account for possible special processing of the edge, such as a thin-film metallic shielding~\cite{AsharpTCS:2025}.

In general, solving the Laplace Equation~(\ref{equ-Laplace}) with the boundary conditions~(\ref{equ-BC2}), (\ref{equ-BC3}), and (\ref{equ-BC4}) is a complex task, particularly when the system lacks cylindrical symmetry due to asymmetric heating sources, such as an \HGthree{} beam. In this work, we adopt a numerical finite element analysis (FEA) approach to solve this problem.

\section{Thermal Analysis \label{sec-thermal}}

\begin{figure*}[t]
    \centering
    \includegraphics[width=1\linewidth]{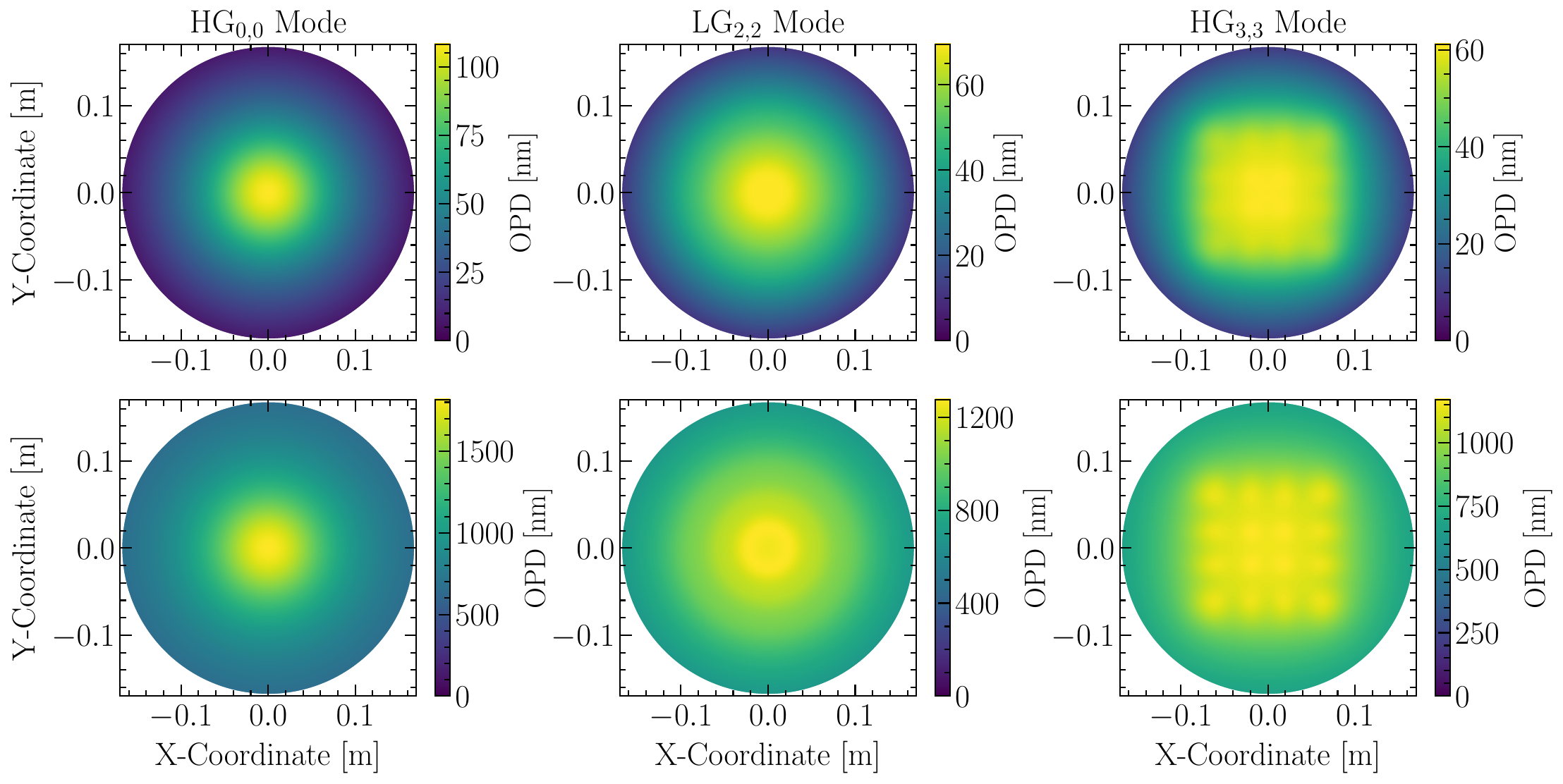}
    \caption{Thermal aberrations of the test mass induced by \HGzero{}, \LGtwo{}, and \HGthree{} absorption from finite element analysis. \textit{Top row:} thermoelastic surface deformation; \textit{Bottom row:} Thermorefractive lensing within the substrate.}
    \label{fig-deformationHOM-2D}
\end{figure*}

\subsection{Thermal distortion mirror maps}

To accurately capture the steady-state thermal response of arbitrary higher-order incident beams, we use the Python package \texttt{FEniCSx} for finite element analysis~\cite{BarattaEtal2023}, utilizing its wrapper \texttt{test-mass-thermal-state} to model the thermal state of test masses under laser heating and thermal actuation~\cite{Cao_TestMassThermalState}. Tab.~\ref{tab-FEA_param} summarizes the key parameters used in the thermal FEA modeling. In this work, we consider an aLIGO-like test mass with a fused silica substrate.

\begin{table}[b]
\centering
\caption{Parameters for the thermal FEA modeling, assuming an aLIGO test mass made of fused silica.}
\begin{tabular}{c|c|c}
\hline \hline 
Parameter & Description & Value \\
\hline 
\textit{R} [m] & Mirror Radius & 0.17  \\ \hline 
\textit{H} [m] & Mirror Height & 0.2 \\ \hline 
\textit{K} [$\mathrm{W/(m\,K)}$] & Thermal Conductivity & 1.38 \\ \hline 
$\epsilon_{\mathrm{f}}$ & Emissivity (Front Surface) & 0.93 \\ \hline 
$\epsilon_{\mathrm{edge}}$ & Emissivity (Edge Surface) & 0.91 \\ \hline 
$T_{0}$ [K] & Ambient Temperature & 293.15 \\ \hline \hline
\end{tabular}
\label{tab-FEA_param}
\end{table}

Fig.~\ref{fig-FEA_3D} illustrates the three-dimensional distribution of the temperature rise (left) and the resulting thermoelastic deformation of the test mass (right). Here, we assume 1~W of \HGthree{} mode absorbed at the HR surface located at $z = 0.1$~m. The temperature distribution is shown for half of the cylindrical test mass, through the $x$--$z$ plane, to reveal the internal temperature gradients within the substrate. As expected, the temperature rise on the front surface is largest at the intensity lobes of the \HGthree{} mode. In addition, the peak temperature rise on the test mass relative to the ambient temperature is 1.83~K for unit-watt \HGthree{} mode absorption and 1.79~K for the \LGtwo{} mode, both significantly smaller than the 3.58~K obtained for the \HGzero{} mode.

The spatially varying temperature profile generates a position-dependent optical path difference (OPD) that acts as a two-dimensional lens for the incident laser beam, altering the beam mode entering the interferometer and causing mode mismatch with the cavity eigenmode. Furthermore, the temperature gradient induces thermal expansion, deforming the test-mass surface as shown on the right side of Fig.~\ref{fig-FEA_3D}. This thermoelastic deformation modifies the HR surface that defines the cavity eigenmode, resulting in thermal aberrations that distort the cavity eigenmode and the laser beam wavefront, thereby degrading their coupling.

\begin{table}[b]
\centering
\caption{Beam size and radius of curvature for a mirror with $R=0.17$~m in an aLIGO-like symmetric cavity of length 3994~m, with clipping loss maintained at 1~ppm.}
\setlength{\tabcolsep}{10pt}
\begin{tabular}{c|c|c|c}
\hline 
\hline 
Mode & \HGzero{} & \LGtwo{} & \HGthree{} \\
\hline  
Beam Size w [cm] & 6.47 & 4.39 & 4.47  \\ \hline 
Mirror RoC [m] & 2052.1 & 2332.8 & 2302.4 \\ \hline \hline 
\end{tabular}
\label{tab-roc_homs}
\end{table}

Fig.~\ref{fig-deformationHOM-2D} shows the thermoelastic distortion ``mirror maps'' on the front surface (top row) and the thermorefractive substrate OPD (bottom row) for unit-Watt \HGzero{}, \LGtwo{}, and \HGthree{} modes incident on and absorbed by the front surface of an aLIGO-like test mass. We consider a symmetric aLIGO-like arm cavity with a cavity length of 3994~m. The mirror radii of curvature and beam sizes have been rescaled according to the parameters in Tab.~\ref{tab-roc_homs}, ensuring that all modes have the same clipping loss of 1~ppm on each test mass. For higher-order incident beams such as the sixth-order \LGtwo{} and \HGthree{} modes, their wider and smoother intensity distributions result in more uniform thermal distortions of smaller amplitude and broader spatial extent, compared to the distortions induced by the fundamental \HGzero{} mode~\cite{Vinet_2007}.

\begin{figure}[t]
    \centering
    \includegraphics[width=1\linewidth]{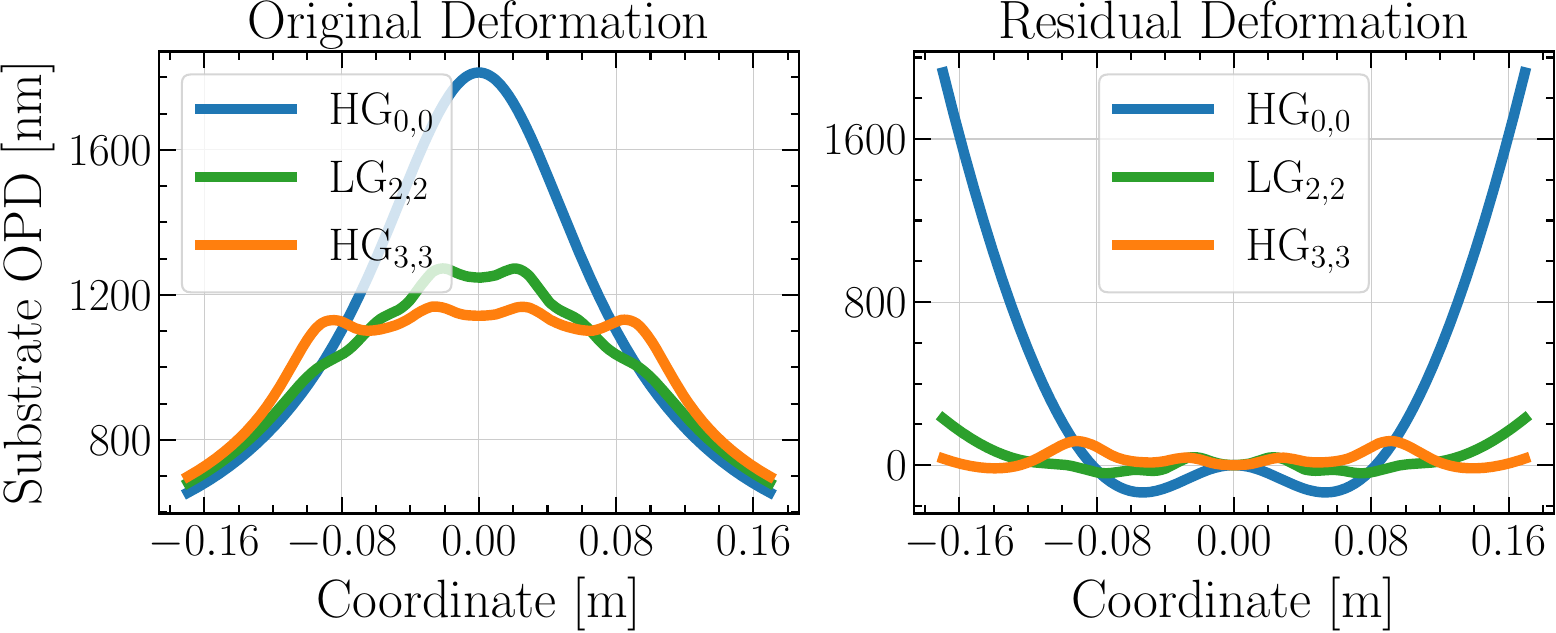}
    \caption{Cross-sectional view of the thermorefractive deformation of the test mass for various incident beams. \textit{Left:} Substrate optical path difference (OPD) without compensation; \textit{Right:} With optimal curvature compensation listed in Tab.~\ref{tab-loss_SB_sub}. The cross section for the \HGthree{} mode is taken along the diagonal, where the OPD is largest due to the brightest intensity lobes.}
    \label{fig-deformationHOM-1D}
\end{figure}

The left panel of Fig.~\ref{fig-deformationHOM-1D} directly compares the thermorefractive OPD for the three cases using a one-dimensional slice through the center of the mirror. For the \HGthree{} mode, the slice is taken along the 45-degree diagonal to account for the lack of cylindrical symmetry, which also corresponds to the line along which the thermal deformation is largest due to the brightest intensity lobes. As shown in the figure, higher-order \LGtwo{} and \HGthree{} exhibit smaller amplitudes and broader spatial distributions than the sharply peaked aberrations produced by the fundamental Gaussian mode, thereby reducing the impact of thermal aberrations on the beam wavefront. For example, the thermal aberration amplitude for the \HGzero{} mode is nearly twice that of the \HGthree{} mode.

\subsection{Intensity-weighted piston and curvature}
We quantify the thermally induced deformation of a cavity mirror as mirror maps, as illustrated in Fig.~\ref{fig-deformationHOM-2D}. For instance, the thermoelastic deformation of the mirror surface can be described by a height map, $\delta h(x,y)$. Upon reflection, this deformation induces a spatially varying optical phase shift on the incident beam.
\begin{equation}
\Phi(x,y) = \frac{4\pi}{\lambda}\,\delta h(x,y),
\label{equ-phasemap}
\end{equation}
where $\lambda$ is the laser wavelength.

\paragraph{Intensity-weighted piston.}
For a given normalized cavity eigenmode $u_{n,m}(x,y)$, with $\int |u_{n,m}(x,y)|^2 \, \mathrm{d}x\,\mathrm{d}y = 1$, the intensity-weighted piston is given as~\cite{Brooks_2021}
\begin{equation}
\begin{aligned}
h^{(n,m)}_{\mathrm{piston}}
&= \bra{u_{n,m}(x,y)} \delta h(x,y) \ket{u_{n,m}(x,y)} \\
&= \int |u_{n,m}(x,y)|^2 \, \delta h(x,y)\, \mathrm{d}x\,\mathrm{d}y .
\end{aligned}
\end{equation}
This term represents a spatially uniform phase shift averaged over the optical mode and is equivalent to an effective cavity length change. It does not modify the cavity eigenmode and is fully compensated by the longitudinal control system. Consequently, it is removed before evaluating higher-order thermal optical effects.

\begin{table}[t]
\centering
\caption{Beam intensity-weighted fit of the thermoelastic mirror surface deformation map, showing piston and the effective curvature.}
\setlength{\tabcolsep}{8pt}
\begin{tabular}{c|c|c}
\hline \hline 
Mode & Piston $h^{(n,m)}_{\mathrm{piston}}$ [nm]&  Curvature $S_{n,m} [\mu D]$ \\
\hline  
\HGzero{} & 85.49 & -3.85 \\ \hline 
\LGtwo{} & 50.39 & -0.82 \\ \hline 
\HGthree{} & 50.84 & -0.37 \\ \hline \hline
\end{tabular}
\label{tab-piston_curv_homs}
\end{table}

\paragraph{Intensity-weighted curvature.}
The thermal aberrations considered here are symmetric with respect to the $X$ and $Y$ axes, due to the fact that the incident beam is centered on the test mass. Consequently, the mirror distortion maps are also symmetric and contain no first-order terms (i.e., tilt). The next higher-order aberration is the curvature.

After subtracting the intensity-weighted piston, the residual deformation is projected onto a quadratic basis to extract the effective curvature seen by the mode. Writing
\begin{equation}
\delta h(x,y) \simeq h^{(n,m)}_{\mathrm{piston}} + S_{n,m} \left(x^2 + y^2\right) .
\end{equation}
The mode-weighted curvature coefficient is obtained from an overlap projection,
\begin{equation}
\begin{aligned}
S_{n,m}
& = \frac{\bra{u_{n,m}(x,y)} \, \delta h(x,y) \cdot (x^2+y^2) \,\ket{u_{n,m}(x,y)}}{\bra{u_{n,m}(x,y)} \, (x^2+y^2)^2 \, \ket{u_{n,m}(x,y)}} \\
& = \frac{
\int |u_{n,m}(x,y)|^2 \, \delta h(x,y) \cdot (x^2+y^2)\,
\mathrm{d}x\,\mathrm{d}y
}{
\int |u_{n,m}(x,y)|^2 \, (x^2+y^2)^2 \,
\mathrm{d}x\,\mathrm{d}y
}.
\end{aligned}
\end{equation}
The corresponding effective change in the mirror radius of curvature is
\begin{equation}
\Delta R_{n,m} = \frac{1}{2S_{n,m}},
\end{equation}
where the sign of $S_{n,m}$ indicates whether the mirror becomes more convex or concave under the respective thermal loading. 

\paragraph{Deviation relative to the fundamental mode.}
To quantify the mitigation of thermal deformation for higher-order modes, we define the effective curvature normalized to the fundamental mode \HGzero{} as
\begin{equation}
\Delta S^{\mathrm{rel}}_{n,m}
=
\frac{ S_{n,m}}{S_{0,0}}.
\end{equation}
This metric directly captures the reduction in thermally induced curvature experienced by a given mode and provides a mode-dependent measure of thermal robustness in high-power interferometric cavities. It also corresponds directly to the power required for thermal actuators, such as quadratic barrel ring heaters (RH), to actively compensate for the curvature change in situ~\cite{Brooks_2016}. Values of $\Delta S^{\mathrm{rel}}_{n,m} < 1$ indicate a reduction in the required curvature correction relative to the fundamental mode. 

For instance, Tab.~\ref{tab-piston_curv_homs} summarizes the beam intensity-weighted fit of the mirror surface deformation maps for all three modes shown in Fig.~\ref{fig-deformationHOM-2D}, including piston and effective curvature. For all three modes, the effective curvature of the mirror surface distortion is negative, indicating that the center of the test mass bulges outward relative to the edges, making the mirror more convex as the effective radius of curvature increases in magnitude, as expected. In addition, compared to the \HGzero{} mode, higher-order modes exhibit substantially smaller curvature changes; for example, the \HGthree{} mode exhibits an effective curvature change reduced by more than an order of magnitude compared to the \HGzero{} mode. This demonstrates that higher-order modes are more thermally robust and lead to significantly less curvature deformation than the fundamental \HGzero{} mode.

\begin{figure}[t]
    \centering
    \includegraphics[width=1\linewidth]{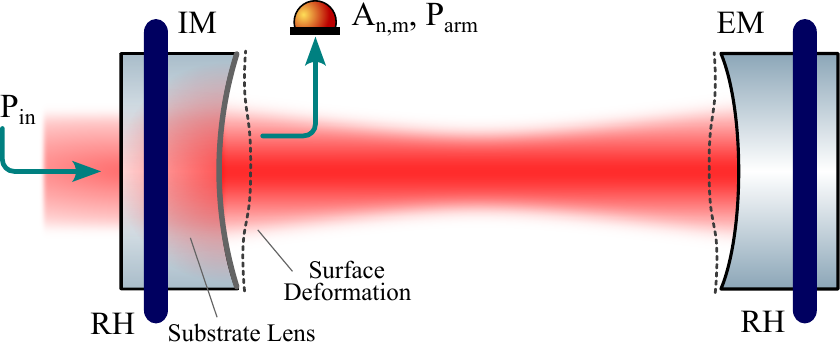}
    \caption{Schematic of an aLIGO-like symmetric cavity. Thermoelastic surface deformation and thermorefractive substrate lensing induced by thermal absorption can deform the cavity eigenmode. Simple quadratic curvature compensation via the barrel ring heater, as already implemented in current detectors, is indicated.}
    \label{fig-FP_cav}
\end{figure}

\section{Mode-Dependent Thermal-Optical Response \label{sec-optical}}
\subsection{Optical layout}

To fully characterize the thermal robustness of higher-order modes relative to the fundamental mode, it is necessary to evaluate the thermally induced optical losses arising from the residual wavefront errors encoded in the thermal maps. For this purpose, we consider an aLIGO-like optical cavity, as illustrated in Fig.~\ref{fig-FP_cav}, with a cavity length of 3994~m. For simplicity, we assume a symmetric cavity in which both the input mirror (IM) and the end mirror (EM) have the same radius of curvature. Because different modes possess distinct intensity distributions, the beam sizes are rescaled to ensure that all modes exhibit the same clipping loss. The resulting beam sizes and corresponding mirror radii of curvature are summarized in Tab.~\ref{tab-roc_homs}, with the clipping loss on the test mass maintained at 1~ppm.

To accurately capture the thermal-optical response, we combine the thermal FEA modeling described above, implemented using \texttt{FEniCSx}, with optical simulations performed using the interferometer simulation software \texttt{Finesse}~\cite{brown_2025_12662017}. We assume that the HR coating of the test mass has a uniform thermal absorption of 0.5~ppm. Thus, for a circulating power of 1~MW, the front surface of the test mass absorbs 0.5~W of optical power.

We first compute the single-bounce scattering loss induced by the unit-Watt thermal distortion maps. Next, the test mass is placed inside the aLIGO-like optical cavity. The simulation procedure is summarized as follows. The three modes (\HGzero{}, \LGtwo{}, and \HGthree{}) are injected into the cavity one at a time. For each mode, we assume the cavity reaches a specified circulating power, and we

\begin{enumerate}
    \vspace{-0.3\baselineskip}
    \item Calculate and rescale the thermal distortion mirror maps for the surface and substrate from \texttt{FEniCSx} according to the absorbed power at the specified circulating power.
    \vspace{-0.3\baselineskip}
    \item Apply the mirror maps to both IM and EM in \texttt{Finesse} and compute the optical gain of the cavity, $\mathrm{OG} = \mathrm{P_{arm} / P_{in}}$, while the cavity is locked to the input mode resonance.
    \vspace{-0.3\baselineskip}
    \item Determine the required input power to achieve the target arm power based on the optical gain.
    \vspace{-0.3\baselineskip}
    \item Compute the corresponding modal impurity using the \texttt{amplitude detector} $\mathrm{A_{n,m}}$ and the \texttt{power detector} $\mathrm{P}_{\mathrm{arm}}$ in \texttt{Finesse}.
    \vspace{-0.3\baselineskip}
\end{enumerate}

\subsection{Single-bounce scattering loss}

\begin{figure}[t]
    \centering
    \includegraphics[width=1\linewidth]{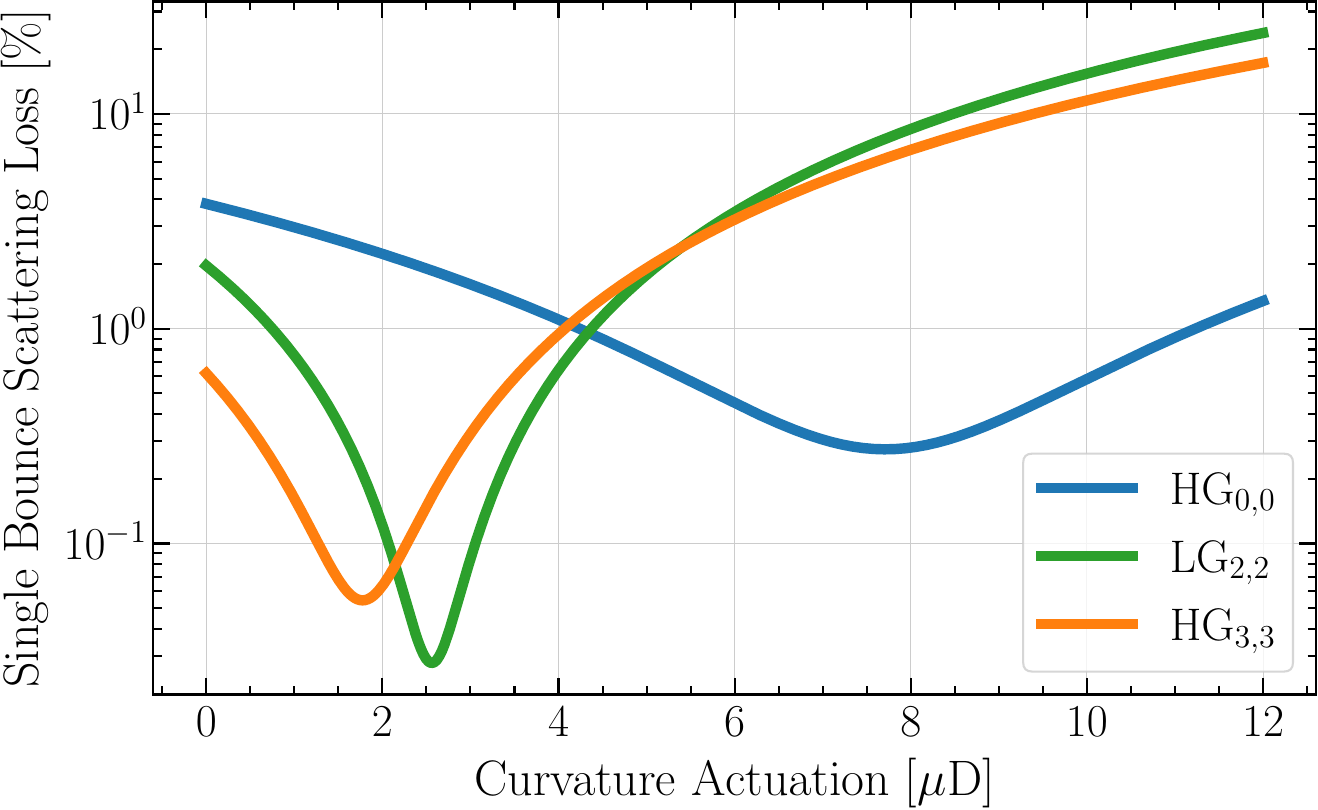}
    \caption{Single-bounce scattering loss induced by the thermally-induced mirror surface distortion map as the applied curvature actuation is increased, shown for the \HGzero{}, \LGtwo{}, and \HGthree{} modes.}
    \label{fig-Scattering_loss_SB_00_22_33}
\end{figure}

To quantify the mode scattering effect of the thermally induced mirror distortion maps on the incident beam, we first compute the single-bounce power loss. The deformed mirror modifies the wavefront of the incident beam. 
Assuming a normalized cavity eigenmode $u_{n,m}(x,y)$, the distorted beam becomes
\begin{equation}
    \tilde{u}_{n,m}(x,y) = u_{n,m}(x,y) \cdot e^{i \Psi(x,y)} ,
\end{equation}
with the mirror phase map $\Psi(x,y)$ given by Eq.~(\ref{equ-phasemap}). The distorted field can be expanded as a combination of cavity eigenmodes. The complex overlap coefficient, $\eta$, between the distorted field and the original field is
\begin{equation}
\begin{aligned}
    \eta &= \Braket{u_{n,m}(x,y) \mid \tilde{u}_{n,m}(x,y)} \\&
    = \int \tilde{u}_{n,m}(x,y) \cdot u_{n,m}^{*}(x,y) \, \mathrm{d}x\,\mathrm{d}y.
\end{aligned}
\end{equation}

The optical scattering loss, $\Gamma$, representing power scattered out of the original eigenmode, is thus
\begin{equation}
    \Gamma = 1 - \lvert \eta \rvert^2 .
\end{equation}

For instance, Fig.~\ref{fig-Scattering_loss_SB_00_22_33} shows the single bounce scattering loss induced by the mirror surface height distortion maps, as the amount of curvature actuation applied to the maps is increased. An optimal curvature correction exists for each of the three modes considered, minimizing the single-bounce scattering loss. 

\begin{table}[t]
\centering
\caption{Single-bounce optical scattering loss from a mirror with a thermoelastically deformed surface, before and after optimal curvature compensation. The required optimal curvature compensation and its value relative to the \HGzero{} mode are shown for comparison.}
\setlength{\tabcolsep}{10pt}
\begin{tabular}{c|c|c|c}
\hline 
\hline 
 & \HGzero{} &  \LGtwo{} & \HGthree{} \\
\hline
$\Gamma^{\text{surf.}}_{1}$ [\%] & 3.8 & 2.0 & 0.62  \\ \hline
$\Gamma^{\text{surf.}}_{2}$ [\%] & 0.27 &  0.028 & 0.054 \\ \hline 
$S_{n,m}^{\mathrm{opt}}$ [$\mu$D] & 7.7 & 2.6 & 1.8 \\ \hline 
$\Delta S^{\mathrm{rel}}_{n,m}$ [\%] & - & 33 & 24 \\ \hline 
\hline
\end{tabular}
\label{tab-loss_SB_surf}
\end{table}

Tab.~\ref{tab-loss_SB_surf} and Tab.~\ref{tab-loss_SB_sub} present the single-bounce power loss $\Gamma^{\mathrm{surf.}}$ induced by mirror surface deformation and the corresponding single-pass power loss $\Gamma^{\mathrm{sub.}}$ incurred during transmission through the substrate lens, respectively. For each table, the first two rows report the power loss computed from the mirror deformation maps shown in Fig.~\ref{fig-deformationHOM-2D} and the corresponding results obtained after applying the optimal curvature correction to remove the quadratic component, respectively. The third row lists the optimal quadratic correction $S_{n,m}^{\mathrm{opt}}$, while the final row gives the required curvature correction relative to the fundamental mode, $\Delta S^{\mathrm{rel}}_{n,m}$. As an example, Tab.~\ref{tab-loss_SB_surf} shows that the \LGtwo{} and \HGthree{} modes require curvature actuation reduced to 33\% and 24\% of the fundamental-mode value, respectively. Consequently, assuming a curvature actuation gain of approximately 1~$\mathrm{\mu D/W}$ for the test mass surface~\cite{Brooks_2016}, the heater power deposited on the barrel of the test mass by the ring heater required to compensate unit-watt absorption is reduced from 7.7~W for the \HGzero{} mode to 2.6~W and 1.8~W for the \LGtwo{} and \HGthree{} modes, respectively. In addition, the power scattering loss due to the residual deformation, $\Gamma^{\mathrm{surf.}}_{2}$, is reduced to approximately one-tenth and one-fifth of the fundamental \HGzero{} mode value.

The right panel of Fig.~\ref{fig-deformationHOM-1D} shows the residual substrate OPD due to the thermorefractive lens for all three modes, after the optimal quadratic corrections listed in Tab.~\ref{tab-loss_SB_sub} are applied to the deformation shown on the left panel. As shown, the much flatter residual deformation for \LGtwo{} and \HGthree{} modes compared to the \HGzero{} mode is the source of the much lower residual scattering loss.

\begin{figure*}[t]
    \centering
    \includegraphics[width=1\linewidth]{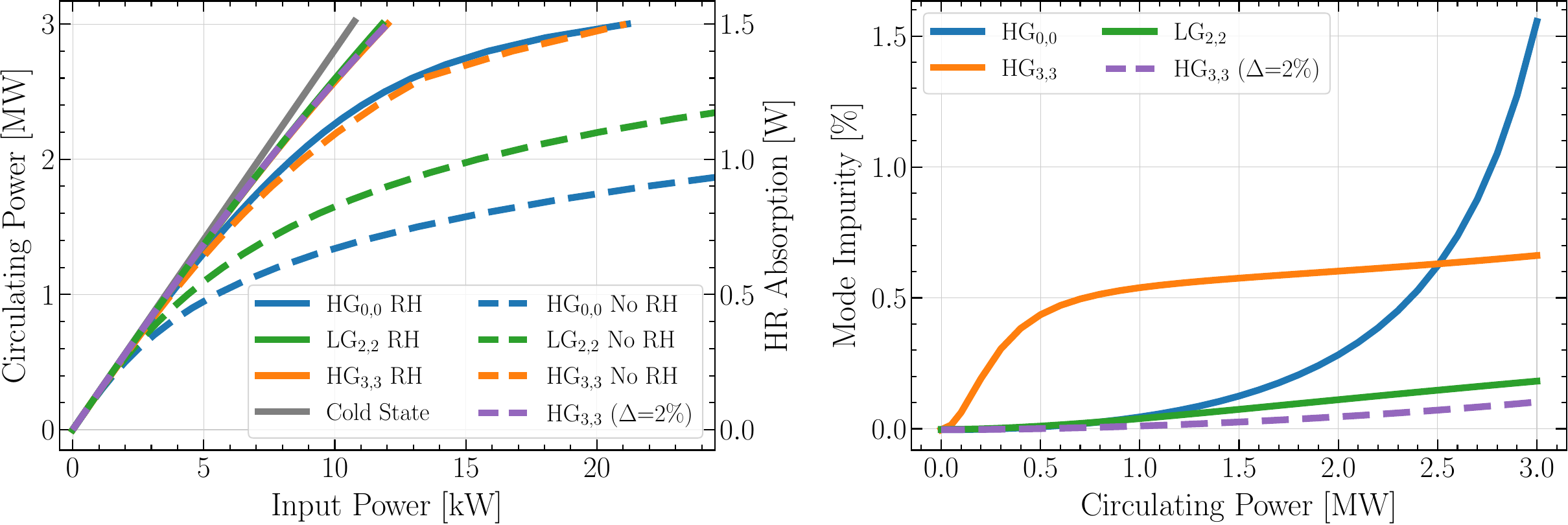}
    \caption{\textit{Left:} Circulating power inside the cavity versus input power. The deviation from the gray line, labeled ``Cold State'' and representing an ideal cavity without thermal aberrations, indicates optical losses due to thermal deformations. Solid lines correspond to results with optimal curvature correction, while dashed lines show results without compensation. \textit{Right:} Modal impurity in the cavity after optimal curvature correction as a function of circulating power. The purple dashed curves correspond to the \HGthree{} mode with 2\% astigmatism.}
    \label{fig-Parm_Pin_00_22_33_impurity}
\end{figure*}

\subsection{Cavity power buildup and modal impurity}

The left panel of Fig.~\ref{fig-Parm_Pin_00_22_33_impurity} shows the circulating power buildup inside the optical cavity as a function of input power for the three modes: \HGzero{} (blue), \LGtwo{} (green), and \HGthree{} (orange). The simulations consider a target circulating power of up to 3~MW, as anticipated for next-generation gravitational-wave detectors~\cite{ETDesignReportUpdate2020}. A secondary $y$-axis, showing the corresponding absorbed absolute power at the front surface of the test mass, is included for easy reference.

The results without any curvature compensation are shown as dashed lines, while the cases with optimal curvature corrections, as listed in Tab.~\ref{tab-loss_SB_surf} and Tab.~\ref{tab-loss_SB_sub}, are shown as solid lines. A gray line labeled ``Cold State'' is included for comparison, representing the ideal cavity coupling with no thermal deformation and constant optical gain. As shown in the figure, applying the corresponding optimal curvature correction leads to higher circulating powers for all three modes at the same input power. In particular, the \LGtwo{} and \HGthree{} modes exhibit greater robustness against thermal aberrations than the fundamental \HGzero{} mode, by maintaining larger optical gains, with the \LGtwo{} mode demonstrating the highest robustness among the three cases.

\begin{table}[t]
\centering
\caption{Single-pass scattering loss due to transmission through a thermorefractive lens in the substrate, before and after optimal curvature compensation. The required optimal curvature compensation and its value relative to the \HGzero{} mode are shown for comparison.}
\setlength{\tabcolsep}{10pt}
\begin{tabular}{c|c|c|c}
\hline 
\hline 
 & \HGzero{} &  \LGtwo{} & \HGthree{} \\
\hline
$\Gamma^{\text{sub.}}_{1}$ [\%] & 80.0 & 46.0 & 14.9  \\ \hline
$\Gamma^{\text{sub.}}_{2}$ [\%] & 10.5 &  1.6 & 4.1 \\ \hline 
$S_{n,m}^{\mathrm{opt}}$ [$\mu$D] & 107.4 & 27.7 & 16.5 \\ \hline 
$\Delta S^{\mathrm{rel}}_{n,m}$ [\%] & - & 26 & 15 \\ \hline 
\hline
\end{tabular}
\label{tab-loss_SB_sub}
\end{table}

\begin{table}[b]
\centering
\caption{Optical gain and modal impurity at 1.5~MW and 3~MW, labeled with subscripts 1 and 2, respectively. For the modal impurity for \HGthree{}, the case with $\Delta=2\%$ astigmatism is also shown. }
\setlength{\tabcolsep}{6pt}
\begin{tabular}{c|c|c|c|c}
\hline\hline
 & \multirow{2}{*}{\HGzero{}} 
 & \multirow{2}{*}{\LGtwo{}} 
 & \multirow{2}{*}{\HGthree{}} 
 & Astigmatic \HGthree{} \\ 
 & & & & ($\Delta=2\%$) \\ 
\hline \hline
OG$_{1}$ & 255.9 & 272.2 & 271.9 & 272.9 \\ \hline
$\Sigma_{1}$ [\%] & 0.12 & 0.074 & 0.58 & 0.026 \\ \hline \hline
OG$_{2}$ & 141.7 & 254.1 & 249.7 & 250.0 \\ \hline
$\Sigma_{2}$ [\%] & 1.6 & 0.18 & 0.66& 0.10 \\ \hline 
\hline
\end{tabular}
\label{tab-OG_homs}
\end{table}

In addition to the cavity power gain, modal impurity provides another commonly used measure of power coupling from the target eigenmode into unwanted co-resonating modes induced by optical aberrations. For an input mode $u_{n,m}(x,y)$ with the cavity locked to the same mode, the modal impurity $\Sigma$ is defined as
\begin{equation}
   \Sigma = 1 - \frac{|\mathrm{A_{n,m}}|^2}{\mathrm{P_{circ}}},
\end{equation}
where $\mathrm{A_{n,m}}$ is the complex amplitude of the input mode inside the cavity, and $\mathrm{P_{circ}}$ is the total circulating power within the cavity.

The right panel of Fig.~\ref{fig-Parm_Pin_00_22_33_impurity} shows the corresponding modal impurity inside the cavity for all three modes. Compared to the fundamental \HGzero{} mode, the higher-order \LGtwo{} mode exhibits much lower modal impurity. This is because the circularly symmetric thermal aberrations generated by \LGtwo{} self-heating scatter power minimally into pseudo-degenerate modes of the same order, consistent with the ``selection rule'' described by Bond et al.~\cite{Bond_2011}. In contrast, the \HGthree{} mode shows significantly higher modal impurity, indicating that a larger fraction of the optical power occupies pseudo-degenerate modes other than the intended cavity eigenmode.

Tab.~\ref{tab-OG_homs} summarizes the optical gain (OG) and modal impurity ($\Sigma$) for all three modes at cavity powers of 1.5~MW and 3~MW, labeled with subscripts 1 and 2, respectively. These correspond to absorbed powers of 0.75~W and 1.5~W for 0.5~ppm of uniform coating absorption. For example, at 3~MW, the optical gains for the \LGtwo{} and \HGthree{} modes are approximately 79\% and 76\% higher than that of the fundamental \HGzero{} mode, respectively. The modal impurity of the \LGtwo{} mode is reduced by roughly a factor of 9 compared to the \HGzero{} mode. However, the modal impurity for \HGthree{} remains relatively large, which can be mitigated by the introduction of astigmatism~\cite{Tao2020}.

\subsection{Modal impurity mitigation with astigmatism}

The circularly asymmetric thermal aberrations induced by \HGthree{} mode heating generate significant modal impurity inside the cavity, due to coupling into pseudo-degenerate modes. This large modal impurity reduces the arm power buildup, which in turn directly diminishes the signal-to-noise ratio, as the gravitational-wave signal scales with the circulating power in the signal-carrying mode within the interferometer arms. Furthermore, large modal impurity can degrade the effective quantum squeezing level at the readout port, potentially introducing a phase mismatch between the squeezed quadrature and the readout field~\cite{McCuller_2021}.

Modal impurity caused by pseudo-degenerate higher-order Hermite-Gaussian modes can be mitigated by introducing astigmatism, either through a static difference in mirror curvature along the two orthogonal directions or via controlled thermal or mechanical actuation~\cite{Tao2020}. In an astigmatic cavity, the round-trip Gouy phase accumulation differs along the $X$ and $Y$ directions. As a result, pseudo-degenerate HG modes of the same order acquire different Gouy phase shifts per round trip, effectively breaking their degeneracy and reducing coupling into unintended modes.

\begin{table}[b]
\centering
\caption{Mirror radii of curvature and beam sizes in the $X$ and $Y$ directions for an \HGthree{} mode with 2\% astigmatism, compared to the non-astigmatic case.}
\begin{tabular}{c|c|c|c|c}
\hline 
\hline 
 $\Delta$ & RoC$_{x}$ [m] &  RoC$_{y}$ [m] & w$_{x}$ [cm] & w$_{y}$ [cm] \\
\hline
0 & 2302.4 & 2302.4 & 4.47 & 4.47  \\ \hline
2\% & 2348.4 & 2256.3 & 4.35 & 4.61  \\ \hline\hline
\end{tabular}
\label{tab-roc_astig}
\end{table}

For an astigmatic cavity mirror, the two radii of curvature can be expressed as
\begin{equation}
\begin{aligned}
    \mathrm{RoC}_{x} &= \mathrm{RoC}_{0} \cdot(1+\Delta) \\
    \mathrm{RoC}_{y} &= \mathrm{RoC}_{0} \cdot(1-\Delta) .
\end{aligned}
\end{equation}
Here, $\mathrm{RoC}_{x}$ and $\mathrm{RoC}_{y}$ denote the radii of curvature in the $X$ and $Y$ directions for the astigmatic mirror, while $\mathrm{RoC}_{0}$ corresponds to the symmetric radius of curvature in the absence of astigmatism. The parameter $\Delta$ represents the magnitude of the astigmatism. For example, Tab.~\ref{tab-roc_astig} lists the radii of curvature for the case $\Delta = 2\%$. Consequently, the beam sizes on the mirror differ in the $X$ and $Y$ directions for an astigmatic cavity.

\begin{figure}[t]
    \centering
    \includegraphics[width=1\linewidth]{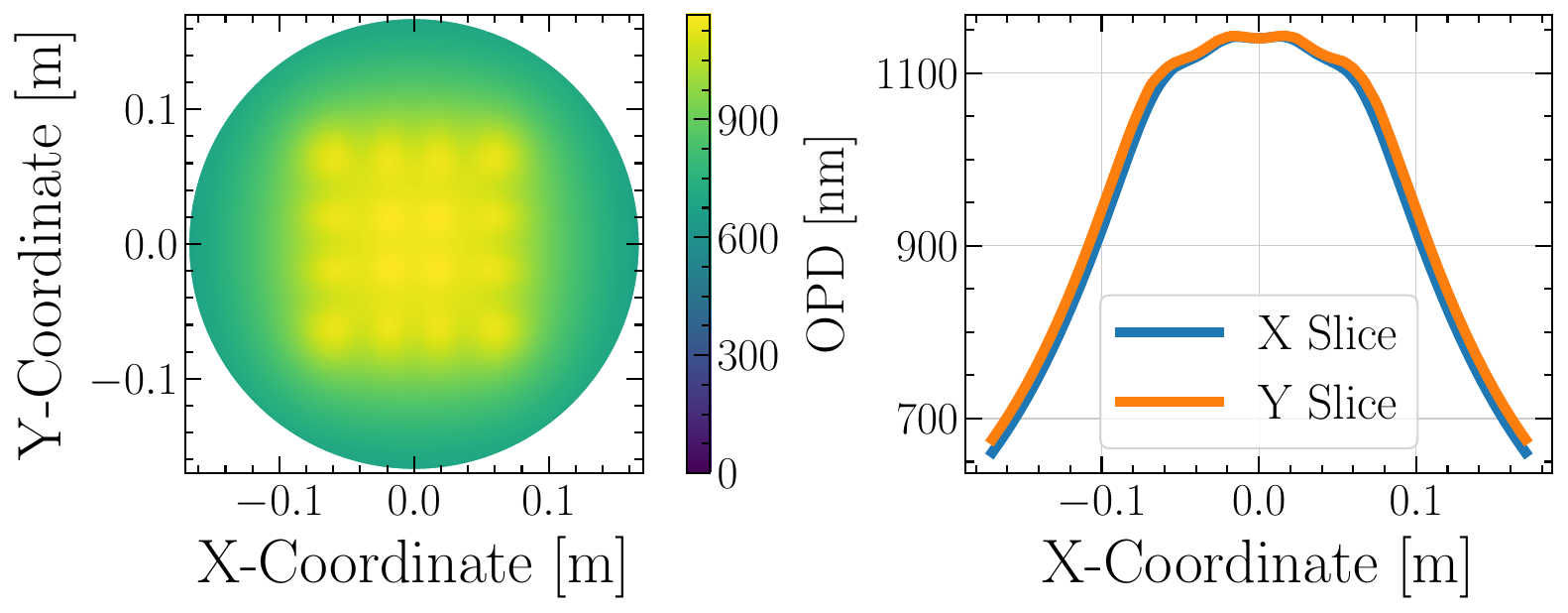}
    \caption{Thermorefractive substrate OPD induced by an $\mathrm{HG}_{3,3}$ mode with 2\% astigmatism, showing a slightly larger beam size along the $Y$ axis compared to $X$.}
    \label{fig-OPD_33_astig}
\end{figure}

Due to the differing beam sizes on the mirror, the thermal response for astigmatic beam heating differs from the non-astigmatic case. Fig.~\ref{fig-OPD_33_astig} shows the substrate OPD for an astigmatic \HGthree{} mode with $\Delta = 2\%$ on the left panel. The right panel presents a cross-sectional view of the OPD along $X$ and $Y$ directions. The thermal deformation varies slightly along the two orthogonal axes.

The cavity power buildup and corresponding modal impurity for the astigmatic \HGthree{} mode ($\Delta = 2\%$) are shown as the purple dashed lines in Fig.~\ref{fig-Parm_Pin_00_22_33_impurity}. The optical gain and modal impurity at 1.5~MW and 3~MW cavity power for the astigmatic \HGthree{} mode are also reported in the last column of Tab.~\ref{tab-OG_homs}. As illustrated, the introduction of astigmatism lifts the problematic pseudo-degeneracy for modes of the same order, enabling the astigmatic \HGthree{} mode to achieve slightly higher cavity power buildup compared to the non-astigmatic case. More importantly, the modal impurity is substantially reduced, resulting in significantly better modal purity than that of the fundamental \HGzero{} mode.

\section{Conclusion and discussion \label{sec-conclusion}}

In this work, we have performed a detailed analysis of the thermal-optical performance of two widely used classes of higher-order modes, specifically, the sixth-order \LGtwo{} and \HGthree{} modes, in comparison to the fundamental \HGzero{} mode. We computed the thermally induced mirror maps using finite-element modeling, capturing both the thermoelastic deformation of the mirror surface and the thermorefractive distortion of the substrate. Due to their more evenly spread-out intensity profiles and the resulting smoother thermal load under absorption, higher-order modes exhibit substantially more uniform thermal distortions than the fundamental mode, with the wavefront distortion significantly reduced.

As a consequence, the required optimal quadratic actuation is reduced to 33\% of that of the fundamental mode for the \LGtwo{} mode and to 24\% for the \HGthree{} mode, indicating that higher-order modes require substantially less actuation power for optimal thermal compensation compared to the fundamental mode. The residual mirror distortion for the \LGtwo{} and \HGthree{} modes results in approximately one order of magnitude and a factor of five times smaller single-bounce scattering loss, respectively, relative to the fundamental mode. This enables significantly improved circulating power buildup and intracavity modal purity in a single aLIGO-like cavity, relative to the currently employed fundamental Gaussian beam. Furthermore, the introduction of small mirror astigmatism has been shown to effectively break the degeneracy among HG modes, substantially enhancing the modal purity for the \HGthree{} mode under self-heating-induced thermal deformation. Consequently, these modes achieve near-optimal performance with only small quadratic actuation, potentially eliminating the need for more complex, higher-resolution wavefront actuations under megawatt-level operation.

This paper highlights the advantages of using flat higher-order laser beams over the sharply peaked fundamental Gaussian mode, particularly in mitigating self-heating-induced thermal aberrations. The resulting reduction in thermal-aberration-induced losses presented in this work represents a significant advantage over the currently used fundamental mode, enabling the potential implementation of higher-order HG modes in high-power laser interferometers such as future gravitational-wave detectors. On the other hand, previous studies have shown that higher-order modes are highly sensitive to realistic surface figure errors present even in state-of-the-art mirrors due to modal degeneracy, particularly in high-finesse optical cavities, which can lead to significant coupling losses~\cite{bond2011, Sorazu_2013, PhysRevD.92.102002}. Several mitigation strategies have been proposed to address this limitation. In particular, the use of corrective mirror coating masks has been investigated as a means to introduce selective losses for the co-resonant degenerate modes and reduce the susceptibility to surface imperfections, thereby improving the robustness and beam quality of higher-order LG modes~\cite{LG06_2026}. In addition, mirror figure errors and modal degeneracy can be made less critical for HG modes, as mirror astigmatism can be introduced to lift the degeneracy~\cite{Tao2020, PhysRevD.103.042008}. A comprehensive assessment is therefore required to combine the advantages of reduced thermal aberrations with the potential improvement in alleviating degeneracy-induced coupling losses, in order to fully evaluate the use of higher-order modes. 

Furthermore, non-uniform coating thermal absorption, such as that arising from localized point defects, can also be analyzed for these candidate higher-order modes. In particular, for Hermite-Gaussian modes, the test mass can always be rotated such that the point defects are positioned at or near intensity nulls (e.g., along the principal axes). In this configuration, thermal absorption and scattering loss induced by point absorbers can be significantly mitigated. These results pave the way for high-power interferometer operation exceeding one megawatt, while mitigating both thermal noise and thermally induced deformations using higher-order laser modes, as anticipated for next-generation gravitational-wave detectors. Future work will investigate the full thermo-optical behavior of higher-order modes in complete gravitational-wave interferometer configurations and experimentally assess their thermo-optical response to realistic imperfections, including non-uniform absorption, beam off-centering on the test mass, and realistic mirror figure errors, as well as implications for control systems and potential parametric instabilities.

\begin{acknowledgments}
The authors thank Matteo Barsuglia for helpful comments during the preparation of this manuscript. Z. Z. acknowledges support from the National Natural Science Foundation of China under Grants Nos. 12433001 and 12021003.

\end{acknowledgments}

\bibliography{references}

@techreport{CEHorizonStudy,
    title={{A Horizon Study for Cosmic Explorer: Science, Observatories, and Community}}, 
    author={M. Evans and R. X. Adhikari and C. Afle and S. W. Ballmer and S. Biscoveanu and S. Borhanian and D. A. Brown and Y. Chen and R. Eisenstein and A. Gruson and A. Gupta and E. D. Hall and R. Huxford and B. Kamai and R. Kashyap and J. S. Kissel and K. Kuns and P. Landry and A. Lenon and G. Lovelace and L. McCuller and K. Ng and A. H. Nitz and J. Read and B. S. Sathyaprakash and D. H. Shoemaker and B. Slagmolen and J. R. Smith and V. Srivastava and L. Sun and S. Vitale and R. Weiss},
    year={2021},
    month = oct,
    Type = {arXiv e-Print},
    Number = {2109.09882},
    eprint={2109.09882},
    archivePrefix={arXiv},
    primaryClass={astro-ph.IM},
    eid = {arXiv:2109.09882},
    url = {https://doi.org/10.48550/arXiv.2109.09882}
}

@techreport{LIGOwhitepaper2025,
    author = {{LIGO Scientific Collaboration}},
    title = {{The LSC Instrument Science White Paper (2025 edition)}},
    Type = {LIGO Technical Report},
    Number = {LIGO-T2400407},
    Year = {2024},
    Month = dec,
    url = {https://dcc.ligo.org/LIGO-T2400407/public}
}

@techreport{PostO5Report:2022,
    Author = {P. Fritschel and K. Kuns and J. Driggers and A. Effler and B. Lantz and D. Ottaway and S. Ballmer and K. Dooley and R. X. Adhikari and M. Evans and B. Farr and G. Gonzalez and P. Schmidt and S. Raja},
    Title = {{Report of the LSC Post-O5 Study Group}},
    Type = {LIGO Technical Report},
    Number = {LIGO-T2200287},
    Year = {2022},
    Month = nov,
    url = {https://dcc.ligo.org/LIGO-T2200287/public}
}

@article{McCuller_2021,
  title = {LIGO's quantum response to squeezed states},
  author = {McCuller, L. and Dwyer, S. E. and Green, A. C. and Yu, Haocun and Kuns, K. and Barsotti, L. and Blair, C. D. and Brown, D. D. and Effler, A. and Evans, M. and Fernandez-Galiana, A. and Fritschel, P. and Frolov, V. V. and Kijbunchoo, N. and Mansell, G. L. and Matichard, F. and Mavalvala, N. and McClelland, D. E. and McRae, T. and Mullavey, A. and Sigg, D. and Slagmolen, B. J. J. and Tse, M. and Vo, T. and Ward, R. L. and Whittle, C. and Abbott, R. and Adams, C. and Adhikari, R. X. and Ananyeva, A. and Appert, S. and Arai, K. and Areeda, J. S. and Asali, Y. and Aston, S. M. and Austin, C. and Baer, A. M. and Ball, M. and Ballmer, S. W. and Banagiri, S. and Barker, D. and Bartlett, J. and Berger, B. K. and Betzwieser, J. and Bhattacharjee, D. and Billingsley, G. and Biscans, S. and Blair, R. M. and Bode, N. and Booker, P. and Bork, R. and Bramley, A. and Brooks, A. F. and Buikema, A. and Cahillane, C. and Cannon, K. C. and Chen, X. and Ciobanu, A. A. and Clara, F. and Compton, C. M. and Cooper, S. J. and Corley, K. R. and Countryman, S. T. and Covas, P. B. and Coyne, D. C. and Datrier, L. E. H. and Davis, D. and Di Fronzo, C. and Dooley, K. L. and Driggers, J. C. and Etzel, T. and Evans, T. M. and Feicht, J. and Fulda, P. and Fyffe, M. and Giaime, J. A. and Giardina, K. D. and Godwin, P. and Goetz, E. and Gras, S. and Gray, C. and Gray, R. and Gustafson, E. K. and Gustafson, R. and Hanks, J. and Hanson, J. and Hardwick, T. and Hasskew, R. K. and Heintze, M. C. and Helmling-Cornell, A. F. and Holland, N. A. and Jones, J. D. and Kandhasamy, S. and Karki, S. and Kasprzack, M. and Kawabe, K. and King, P. J. and Kissel, J. S. and Kumar, Rahul and Landry, M. and Lane, B. B. and Lantz, B. and Laxen, M. and Lecoeuche, Y. K. and Leviton, J. and Liu, J. and Lormand, M. and Lundgren, A. P. and Macas, R. and MacInnis, M. and Macleod, D. M. and M\'arka, S. and M\'arka, Z. and Martynov, D. V. and Mason, K. and Massinger, T. J. and McCarthy, R. and McCormick, S. and McIver, J. and Mendell, G. and Merfeld, K. and Merilh, E. L. and Meylahn, F. and Mistry, T. and Mittleman, R. and Moreno, G. and Mow-Lowry, C. M. and Mozzon, S. and Nelson, T. J. N. and Nguyen, P. and Nuttall, L. K. and Oberling, J. and Oram, Richard J. and Osthelder, C. and Ottaway, D. J. and Overmier, H. and Palamos, J. R. and Parker, W. and Payne, E. and Pele, A. and Penhorwood, R. and Perez, C. J. and Pirello, M. and Radkins, H. and Ramirez, K. E. and Richardson, J. W. and Riles, K. and Robertson, N. A. and Rollins, J. G. and Romel, C. L. and Romie, J. H. and Ross, M. P. and Ryan, K. and Sadecki, T. and Sanchez, E. J. and Sanchez, L. E. and Saravanan, T. R. and Savage, R. L. and Schaetzl, D. and Schnabel, R. and Schofield, R. M. S. and Schwartz, E. and Sellers, D. and Shaffer, T. and Smith, J. R. and Soni, S. and Sorazu, B. and Spencer, A. P. and Strain, K. A. and Sun, L. and Szczepa\ifmmode \acute{n}\else \'{n}\fi{}czyk, M. J. and Thomas, M. and Thomas, P. and Thorne, K. A. and Toland, K. and Torrie, C. I. and Traylor, G. and Urban, A. L. and Vajente, G. and Valdes, G. and Vander-Hyde, D. C. and Veitch, P. J. and Venkateswara, K. and Venugopalan, G. and Viets, A. D. and Vorvick, C. and Wade, M. and Warner, J. and Weaver, B. and Weiss, R. and Willke, B. and Wipf, C. C. and Xiao, L. and Yamamoto, H. and Yu, Hang and Zhang, L. and Zucker, M. E. and Zweizig, J.},
  journal = {Phys. Rev. D},
  volume = {104},
  issue = {6},
  pages = {062006},
  numpages = {29},
  year = {2021},
  month = {Sep},
  publisher = {American Physical Society},
  doi = {10.1103/PhysRevD.104.062006},
  url = {https://link.aps.org/doi/10.1103/PhysRevD.104.062006}
}

@article{Tao2020,
  title = {Higher-order Hermite-Gauss modes as a robust flat beam in interferometric gravitational wave detectors},
  author = {Tao, Liu and Green, Anna and Fulda, Paul},
  journal = {Phys. Rev. D},
  volume = {102},
  issue = {12},
  pages = {122002},
  numpages = {10},
  year = {2020},
  month = {Dec},
  publisher = {American Physical Society},
  doi = {10.1103/PhysRevD.102.122002},
  url = {https://link.aps.org/doi/10.1103/PhysRevD.102.122002}
}

@article{Bond_2011,
   title={Higher order Laguerre-Gauss mode degeneracy in realistic, high finesse cavities},
   volume={84},
   ISSN={1550-2368},
   url={http://dx.doi.org/10.1103/PhysRevD.84.102002},
   DOI={10.1103/physrevd.84.102002},
   number={10},
   journal={Physical Review D},
   publisher={American Physical Society (APS)},
   author={Bond, Charlotte and Fulda, Paul and Carbone, Ludovico and Kokeyama, Keiko and Freise, Andreas},
   year={2011},
   month=nov }

@article{Capote_2025,
   title={Advanced LIGO detector performance in the fourth observing run},
   volume={111},
   ISSN={2470-0029},
   url={http://dx.doi.org/10.1103/PhysRevD.111.062002},
   DOI={10.1103/physrevd.111.062002},
   number={6},
   journal={Physical Review D},
   publisher={American Physical Society (APS)},
   author={Capote, E. and Jia, W. and Aritomi, N. and Nakano, M. and Xu, V. and Abbott, R. and Abouelfettouh, I. and Adhikari, R. X. and Ananyeva, A. and Appert, S. and Apple, S. K. and Arai, K. and Aston, S. M. and Ball, M. and Ballmer, S. W. and Barker, D. and Barsotti, L. and Berger, B. K. and Betzwieser, J. and Bhattacharjee, D. and Billingsley, G. and Biscans, S. and Blair, C. D. and Bode, N. and Bonilla, E. and Bossilkov, V. and Branch, A. and Brooks, A. F. and Brown, D. D. and Bryant, J. and Cahillane, C. and Cao, H. and Clara, F. and Collins, J. and Compton, C. M. and Cottingham, R. and Coyne, D. C. and Crouch, R. and Csizmazia, J. and Cumming, A. and Dartez, L. P. and Davis, D. and Demos, N. and Dohmen, E. and Driggers, J. C. and Dwyer, S. E. and Effler, A. and Ejlli, A. and Etzel, T. and Evans, M. and Feicht, J. and Frey, R. and Frischhertz, W. and Fritschel, P. and Frolov, V. V. and Fuentes-Garcia, M. and Fulda, P. and Fyffe, M. and Ganapathy, D. and Gateley, B. and Gayer, T. and Giaime, J. A. and Giardina, K. D. and Glanzer, J. and Goetz, E. and Goetz, R. and Goodwin-Jones, A. W. and Gras, S. and Gray, C. and Griffith, D. and Grote, H. and Guidry, T. and Gurs, J. and Hall, E. D. and Hanks, J. and Hanson, J. and Heintze, M. C. and Helmling-Cornell, A. F. and Holland, N. A. and Hoyland, D. and Huang, H. Y. and Inoue, Y. and James, A. L. and Jamies, A. and Jennings, A. and Jones, D. H. and Kabagoz, H. B. and Karat, S. and Karki, S. and Kasprzack, M. and Kawabe, K. and Kijbunchoo, N. and King, P. J. and Kissel, J. S. and Komori, K. and Kontos, A. and Kumar, Rahul and Kuns, K. and Landry, M. and Lantz, B. and Laxen, M. and Lee, K. and Lesovsky, M. and Villarreal, F. Llamas and Lormand, M. and Loughlin, H. A. and Macas, R. and MacInnis, M. and Makarem, C. N. and Mannix, B. and Mansell, G. L. and Martin, R. M. and Mason, K. and Matichard, F. and Mavalvala, N. and Maxwell, N. and McCarrol, G. and McCarthy, R. and McClelland, D. E. and McCormick, S. and McRae, T. and Mera, F. and Merilh, E. L. and Meylahn, F. and Mittleman, R. and Moraru, D. and Moreno, G. and Mullavey, A. and Nelson, T. J. N. and Neunzert, A. and Notte, J. and Oberling, J. and O’Hanlon, T. and Osthelder, C. and Ottaway, D. J. and Overmier, H. and Parker, W. and Patane, O. and Pele, A. and Pham, H. and Pirello, M. and Pullin, J. and Quetschke, V. and Ramirez, K. E. and Ransom, K. and Reyes, J. and Richardson, J. W. and Robinson, M. and Rollins, J. G. and Romel, C. L. and Romie, J. H. and Ross, M. P. and Ryan, K. and Sadecki, T. and Sanchez, A. and Sanchez, E. J. and Sanchez, L. E. and Savage, R. L. and Schaetzl, D. and Schiworski, M. G. and Schnabel, R. and Schofield, R. M. S. and Schwartz, E. and Sellers, D. and Shaffer, T. and Short, R. W. and Sigg, D. and Slagmolen, B. J. J. and Soike, C. and Soni, S. and Srivastava, V. and Sun, L. and Tanner, D. B. and Thomas, M. and Thomas, P. and Thorne, K. A. and Todd, M. R. and Torrie, C. I. and Traylor, G. and Ubhi, A. S. and Vajente, G. and Vanosky, J. and Vecchio, A. and Veitch, P. J. and Vibhute, A. M. and von Reis, E. R. G. and Warner, J. and Weaver, B. and Weiss, R. and Whittle, C. and Willke, B. and Wipf, C. C. and Wright, J. L. and Yamamoto, H. and Zhang, L. and Zucker, M. E.},
   year={2025},
   month=mar 
}

@article{Acernese_2015,
    doi = {10.1088/0264-9381/32/2/024001},
    url = {https://doi.org/10.1088/0264-9381/32/2/024001},
    year = {2014},
    month = {dec},
    publisher = {IOP Publishing},
    volume = {32},
    number = {2},
    pages = {024001},
    author = {Acernese, F and Agathos, M and Agatsuma, K and Aisa, D and Allemandou, N and Allocca, A and Amarni, J and Astone, P and Balestri, G and Ballardin, G and Barone, F and Baronick, J-P and Barsuglia, M and Basti, A and Basti, F and Bauer, Th S and Bavigadda, V and Bejger, M and Beker, M G and Belczynski, C and Bersanetti, D and Bertolini, A and Bitossi, M and Bizouard, M A and Bloemen, S and Blom, M and Boer, M and Bogaert, G and Bondi, D and Bondu, F and Bonelli, L and Bonnand, R and Boschi, V and Bosi, L and Bouedo, T and Bradaschia, C and Branchesi, M and Briant, T and Brillet, A and Brisson, V and Bulik, T and Bulten, H J and Buskulic, D and Buy, C and Cagnoli, G and Calloni, E and Campeggi, C and Canuel, B and Carbognani, F and Cavalier, F and Cavalieri, R and Cella, G and Cesarini, E and Mottin, E Chassande- and Chincarini, A and Chiummo, A and Chua, S and Cleva, F and Coccia, E and Cohadon, P-F and Colla, A and Colombini, M and Conte, A and Coulon, J-P and Cuoco, E and Dalmaz, A and D’Antonio, S and Dattilo, V and Davier, M and Day, R and Debreczeni, G and Degallaix, J and Deléglise, S and Pozzo, W Del and Dereli, H and Rosa, R De and Fiore, L Di and Lieto, A Di and Virgilio, A Di and Doets, M and Dolique, V and Drago, M and Ducrot, M and Endrőczi, G and Fafone, V and Farinon, S and Ferrante, I and Ferrini, F and Fidecaro, F and Fiori, I and Flaminio, R and Fournier, J-D and Franco, S and Frasca, S and Frasconi, F and Gammaitoni, L and Garufi, F and Gaspard, M and Gatto, A and Gemme, G and Gendre, B and Genin, E and Gennai, A and Ghosh, S and Giacobone, L and Giazotto, A and Gouaty, R and Granata, M and Greco, G and Groot, P and Guidi, G M and Harms, J and Heidmann, A and Heitmann, H and Hello, P and Hemming, G and Hennes, E and Hofman, D and Jaranowski, P and Jonker, R J G and Kasprzack, M and Kéfélian, F and Kowalska, I and Kraan, M and Królak, A and Kutynia, A and Lazzaro, C and Leonardi, M and Leroy, N and Letendre, N and Li, T G F and Lieunard, B and Lorenzini, M and Loriette, V and Losurdo, G and Magazzù, C and Majorana, E and Maksimovic, I and Malvezzi, V and Man, N and Mangano, V and Mantovani, M and Marchesoni, F and Marion, F and Marque, J and Martelli, F and Martellini, L and Masserot, A and Meacher, D and Meidam, J and Mezzani, F and Michel, C and Milano, L and Minenkov, Y and Moggi, A and Mohan, M and Montani, M and Morgado, N and Mours, B and Mul, F and Nagy, M F and Nardecchia, I and Naticchioni, L and Nelemans, G and Neri, I and Neri, M and Nocera, F and Pacaud, E and Palomba, C and Paoletti, F and Paoli, A and Pasqualetti, A and Passaquieti, R and Passuello, D and Perciballi, M and Petit, S and Pichot, M and Piergiovanni, F and Pillant, G and Piluso, A and Pinard, L and Poggiani, R and Prijatelj, M and Prodi, G A and Punturo, M and Puppo, P and Rabeling, D S and Rácz, I and Rapagnani, P and Razzano, M and Re, V and Regimbau, T and Ricci, F and Robinet, F and Rocchi, A and Rolland, L and Romano, R and Rosińska, D and Ruggi, P and Saracco, E and Sassolas, B and Schimmel, F and Sentenac, D and Sequino, V and Shah, S and Siellez, K and Straniero, N and Swinkels, B and Tacca, M and Tonelli, M and Travasso, F and Turconi, M and Vajente, G and van Bakel, N and van Beuzekom, M and van den Brand, J F J and Van Den Broeck, C and van der Sluys, M V and van Heijningen, J and Vasúth, M and Vedovato, G and Veitch, J and Verkindt, D and Vetrano, F and Viceré, A and Vinet, J-Y and Visser, G and Vocca, H and Ward, R and Was, M and Wei, L-W and Yvert, M and żny, A Zadro and Zendri, J-P},
    title = {Advanced Virgo: a second-generation interferometric gravitational wave detector},
    journal = {Classical and Quantum Gravity}
}

@article{PhysRevLett.127.071101,
  title = {Low Mechanical Loss ${\mathrm{TiO}}_{2}:{\mathrm{GeO}}_{2}$ Coatings for Reduced Thermal Noise in Gravitational Wave Interferometers},
  author = {Vajente, Gabriele and Yang, Le and Davenport, Aaron and Fazio, Mariana and Ananyeva, Alena and Zhang, Liyuan and Billingsley, Garilynn and Prasai, Kiran and Markosyan, Ashot and Bassiri, Riccardo and Fejer, Martin M. and Chicoine, Martin and Schiettekatte, Fran\ifmmode \mbox{\c{c}}\else \c{c}\fi{}ois and Menoni, Carmen S.},
  journal = {Phys. Rev. Lett.},
  volume = {127},
  issue = {7},
  pages = {071101},
  numpages = {7},
  year = {2021},
  month = {Aug},
  publisher = {American Physical Society},
  doi = {10.1103/PhysRevLett.127.071101},
  url = {https://link.aps.org/doi/10.1103/PhysRevLett.127.071101}
}

@article{Adhikari_2020,
doi = {10.1088/1361-6382/ab9143},
url = {https://doi.org/10.1088/1361-6382/ab9143},
year = {2020},
month = {jul},
publisher = {IOP Publishing},
volume = {37},
number = {16},
pages = {165003},
author = {Adhikari, R X and Arai, K and Brooks, A F and Wipf, C and Aguiar, O and Altin, P and Barr, B and Barsotti, L and Bassiri, R and Bell, A and Billingsley, G and Birney, R and Blair, D and Bonilla, E and Briggs, J and Brown, D D and Byer, R and Cao, H and Constancio, M and Cooper, S and Corbitt, T and Coyne, D and Cumming, A and Daw, E and deRosa, R and Eddolls, G and Eichholz, J and Evans, M and Fejer, M and Ferreira, E C and Freise, A and Frolov, V V and Gras, S and Green, A and Grote, H and Gustafson, E and Hall, E D and Hammond, G and Harms, J and Harry, G and Haughian, K and Heinert, D and Heintze, M and Hellman, F and Hennig, J and Hennig, M and Hild, S and Hough, J and Johnson, W and Kamai, B and Kapasi, D and Komori, K and Koptsov, D and Korobko, M and Korth, W Z and Kuns, K and Lantz, B and Leavey, S and Magana-Sandoval, F and Mansell, G and Markosyan, A and Markowitz, A and Martin, I and Martin, R and Martynov, D and McClelland, D E and McGhee, G and McRae, T and Mills, J and Mitrofanov, V and Molina-Ruiz, M and Mow-Lowry, C and Munch, J and Murray, P and Ng, S and Okada, M A and Ottaway, D J and Prokhorov, L and Quetschke, V and Reid, S and Reitze, D and Richardson, J and Robie, R and Romero-Shaw, I and Route, R and Rowan, S and Schnabel, R and Schneewind, M and Seifert, F and Shaddock, D and Shapiro, B and Shoemaker, D and Silva, A S and Slagmolen, B and Smith, J and Smith, N and Steinlechner, J and Strain, K and Taira, D and Tait, S and Tanner, D and Tornasi, Z and Torrie, C and Van Veggel, M and Vanheijningen, J and Veitch, P and Wade, A and Wallace, G and Ward, R and Weiss, R and Wessels, P and Willke, B and Yamamoto, H and Yap, M J and Zhao, C},
title = {A cryogenic silicon interferometer for gravitational-wave detection},
journal = {Classical and Quantum Gravity}
}

@article{Mours_2006,
doi = {10.1088/0264-9381/23/20/001},
url = {https://doi.org/10.1088/0264-9381/23/20/001},
year = {2006},
month = {sep},
publisher = {},
volume = {23},
number = {20},
pages = {5777},
author = {Mours, Benoît and Tournefier, Edwige and Vinet, Jean-Yves},
title = {Thermal noise reduction in interferometric gravitational wave antennas: using high order TEM modes},
journal = {Classical and Quantum Gravity}
}

@article{PhysRevD.82.042003,
  title = {Thermal noise in advanced gravitational wave interferometric antennas: A comparison between arbitrary order Hermite and Laguerre Gaussian modes},
  author = {Vinet, Jean-Yves},
  journal = {Phys. Rev. D},
  volume = {82},
  issue = {4},
  pages = {042003},
  numpages = {9},
  year = {2010},
  month = {Aug},
  publisher = {American Physical Society},
  doi = {10.1103/PhysRevD.82.042003},
  url = {https://link.aps.org/doi/10.1103/PhysRevD.82.042003}
}

@article{PhysRevD.82.012002,
  title = {Experimental demonstration of higher-order Laguerre-Gauss mode interferometry},
  author = {Fulda, Paul and Kokeyama, Keiko and Chelkowski, Simon and Freise, Andreas},
  journal = {Phys. Rev. D},
  volume = {82},
  issue = {1},
  pages = {012002},
  numpages = {7},
  year = {2010},
  month = {Jul},
  publisher = {American Physical Society},
  doi = {10.1103/PhysRevD.82.012002},
  url = {https://link.aps.org/doi/10.1103/PhysRevD.82.012002}
}

@article{bond2011,
  title = {Higher order Laguerre-Gauss mode degeneracy in realistic, high finesse cavities},
  author = {Bond, Charlotte and Fulda, Paul and Carbone, Ludovico and Kokeyama, Keiko and Freise, Andreas},
  journal = {Phys. Rev. D},
  volume = {84},
  issue = {10},
  pages = {102002},
  numpages = {12},
  year = {2011},
  month = {Nov},
  publisher = {American Physical Society},
  doi = {10.1103/PhysRevD.84.102002},
  url = {https://link.aps.org/doi/10.1103/PhysRevD.84.102002}
}

@article{Sorazu_2013,
doi = {10.1088/0264-9381/30/3/035004},
url = {https://doi.org/10.1088/0264-9381/30/3/035004},
year = {2013},
month = {jan},
publisher = {IOP Publishing},
volume = {30},
number = {3},
pages = {035004},
author = {Sorazu, B and Fulda, P J and Barr, B W and Bell, A S and Bond, C and Carbone, L and Freise, A and Hild, S and Huttner, S H and Macarthur, J and Strain, K A},
title = {Experimental test of higher-order Laguerre–Gauss modes in the 10 m Glasgow prototype interferometer},
journal = {Classical and Quantum Gravity}
}

@article{PhysRevD.103.042008,
  title = {Higher-order Hermite-Gauss modes for gravitational waves detection},
  author = {Ast, Stefan and Di Pace, Sibilla and Millo, Jacques and Pichot, Mikha\"el and Turconi, Margherita and Christensen, Nelson and Chaibi, Walid},
  journal = {Phys. Rev. D},
  volume = {103},
  issue = {4},
  pages = {042008},
  numpages = {17},
  year = {2021},
  month = {Feb},
  publisher = {American Physical Society},
  doi = {10.1103/PhysRevD.103.042008},
  url = {https://link.aps.org/doi/10.1103/PhysRevD.103.042008}
}

@article{PhysRevLett.132.101402,
  title = {Experimental Demonstrations of Alignment and Mode Matching in Optical Cavities with Higher-Order Hermite-Gauss Modes},
  author = {Tao, Liu and Fulda, Paul},
  journal = {Phys. Rev. Lett.},
  volume = {132},
  issue = {10},
  pages = {101402},
  numpages = {6},
  year = {2024},
  month = {Mar},
  publisher = {American Physical Society},
  doi = {10.1103/PhysRevLett.132.101402},
  url = {https://link.aps.org/doi/10.1103/PhysRevLett.132.101402}
}

@article{10.1063/5.0137085,
    author = {Behren, B. von and Heinze, Joscha and Bode, Nina and Willke, Benno},
    title = {High-power laser beam in higher-order Hermite–Gaussian modes},
    journal = {Applied Physics Letters},
    volume = {122},
    number = {19},
    pages = {191105},
    year = {2023},
    month = {05},
    issn = {0003-6951},
    doi = {10.1063/5.0137085},
    url = {https://doi.org/10.1063/5.0137085},
}

@article{PhysRevLett.128.083606,
  title = {Observation of Squeezed States of Light in Higher-Order Hermite-Gaussian Modes with a Quantum Noise Reduction of up to 10 dB},
  author = {Heinze, Joscha and Willke, Benno and Vahlbruch, Henning},
  journal = {Phys. Rev. Lett.},
  volume = {128},
  issue = {8},
  pages = {083606},
  numpages = {6},
  year = {2022},
  month = {Feb},
  publisher = {American Physical Society},
  doi = {10.1103/PhysRevLett.128.083606},
  url = {https://link.aps.org/doi/10.1103/PhysRevLett.128.083606}
}

@article{PhysRevLett.129.031101,
  title = {10 dB Quantum-Enhanced Michelson Interferometer with Balanced Homodyne Detection},
  author = {Heinze, Joscha and Danzmann, Karsten and Willke, Benno and Vahlbruch, Henning},
  journal = {Phys. Rev. Lett.},
  volume = {129},
  issue = {3},
  pages = {031101},
  numpages = {6},
  year = {2022},
  month = {Jul},
  publisher = {American Physical Society},
  doi = {10.1103/PhysRevLett.129.031101},
  url = {https://link.aps.org/doi/10.1103/PhysRevLett.129.031101}
}

@article{Lee2019,
author = {Lee, James and Alexander, S. and Kevan, S. and Roy, Suvayan and McMorran, Benjamin},
year = {2019},
month = {03},
pages = {},
title = {Laguerre–Gauss and Hermite–Gauss soft X-ray states generated using diffractive optics},
volume = {13},
journal = {Nature Photonics},
doi = {10.1038/s41566-018-0328-8}
}

@article{PhysRevLett.110.043601,
  title = {Object Identification Using Correlated Orbital Angular Momentum States},
  author = {Uribe-Patarroyo, N\'estor and Fraine, Andrew and Simon, David S. and Minaeva, Olga and Sergienko, Alexander V.},
  journal = {Phys. Rev. Lett.},
  volume = {110},
  issue = {4},
  pages = {043601},
  numpages = {5},
  year = {2013},
  month = {Jan},
  publisher = {American Physical Society},
  doi = {10.1103/PhysRevLett.110.043601},
  url = {https://link.aps.org/doi/10.1103/PhysRevLett.110.043601}
}

@article{10.1063/1.4869819,
    author = {Sun, Hengxin and Liu, Kui and Liu, Zunlong and Guo, Pengliang and Zhang, Junxiang and Gao, Jiangrui},
    title = {Small-displacement measurements using high-order Hermite-Gauss modes},
    journal = {Applied Physics Letters},
    volume = {104},
    number = {12},
    pages = {121908},
    year = {2014},
    month = {03},
    issn = {0003-6951},
    doi = {10.1063/1.4869819},
    url = {https://doi.org/10.1063/1.4869819},
}

@article{PhysRevLett.98.083602,
  title = {Tools for Multimode Quantum Information: Modulation, Detection, and Spatial Quantum Correlations},
  author = {Lassen, M. and Delaubert, V. and Janousek, J. and Wagner, K. and Bachor, H.-A. and Lam, P. K. and Treps, N. and Buchhave, P. and Fabre, C. and Harb, C. C.},
  journal = {Phys. Rev. Lett.},
  volume = {98},
  issue = {8},
  pages = {083602},
  numpages = {4},
  year = {2007},
  month = {Feb},
  publisher = {American Physical Society},
  doi = {10.1103/PhysRevLett.98.083602},
  url = {https://link.aps.org/doi/10.1103/PhysRevLett.98.083602}
}

@article{Brooks_2016,
   title={Overview of Advanced LIGO adaptive optics},
   volume={55},
   ISSN={1539-4522},
   url={http://dx.doi.org/10.1364/AO.55.008256},
   DOI={10.1364/ao.55.008256},
   number={29},
   journal={Applied Optics},
   publisher={Optica Publishing Group},
   author={Brooks, Aidan F. and Abbott, Benjamin and Arain, Muzammil A. and Ciani, Giacomo and Cole, Ayodele and Grabeel, Greg and Gustafson, Eric and Guido, Chris and Heintze, Matthew and Heptonstall, Alastair and Jacobson, Mindy and Kim, Won and King, Eleanor and Lynch, Alexander and O’Connor, Stephen and Ottaway, David and Mailand, Ken and Mueller, Guido and Munch, Jesper and Sannibale, Virginio and Shao, Zhenhua and Smith, Michael and Veitch, Peter and Vo, Thomas and Vorvick, Cheryl and Willems, Phil},
   year={2016},
   month=oct, pages={8256} }

@article{Vinet_2007,
doi = {10.1088/0264-9381/24/15/008},
url = {https://doi.org/10.1088/0264-9381/24/15/008},
year = {2007},
month = {jul},
publisher = {},
volume = {24},
number = {15},
pages = {3897},
author = {Vinet, Jean-Yves},
title = {Reducing thermal effects in mirrors of advanced gravitational wave interferometric detectors},
journal = {Classical and Quantum Gravity}
}

@article{Vinet2009zz,
    author = "Vinet, Jean-Yves",
    title = "{On special optical modes and thermal issues in advanced gravitational wave interferometric detectors}",
    doi = "10.12942/lrr-2009-5",
    journal = "Living Rev. Rel.",
    volume = "12",
    pages = "5",
    year = "2009"
}

@misc{BarattaEtal2023,
  title     = {{DOLFINx}: the next generation {FEniCS} problem solving environment},
  author    = {Baratta, Igor A. and Dean, Joseph P. and Dokken, J{\o}rgen S. and Habera, Michal and Hale, Jack S. and Richardson, Chris N. and Rognes, Marie E. and Scroggs, Matthew W. and Sime, Nathan and Wells, Garth N.},
  doi       = {10.5281/zenodo.10447666},
  year      = {2023},
  howpublished = {preprint}
}

@software{Cao_TestMassThermalState,
  author       = {Cao, Huy Tuong and Brown, Daniel and Smith, Rory and Brooks, Aidan},
  title        = {Thermal State Modelling of Optical Elements},
  howpublished = {\url{https://gitlab.com/ifosim/test-mass-thermal-state}},
  note         = {GPL v3 licensed software. Accessed: 26 January 2026}
}

@article{Brooks_2021,
   title={Point absorbers in Advanced LIGO},
   volume={60},
   ISSN={2155-3165},
   url={http://dx.doi.org/10.1364/AO.419689},
   DOI={10.1364/ao.419689},
   number={13},
   journal={Applied Optics},
   publisher={Optica Publishing Group},
   author={Brooks, Aidan F. and Vajente, Gabriele and Yamamoto, Hiro and Abbott, Rich and Adams, Carl and Adhikari, Rana X. and Ananyeva, Alena and Appert, Stephen and Arai, Koji and Areeda, Joseph S. and Asali, Yasmeen and Aston, Stuart M. and Austin, Corey and Baer, Anne M. and Ball, Matthew and Ballmer, Stefan W. and Banagiri, Sharan and Barker, David and Barsotti, Lisa and Bartlett, Jeffrey and Berger, Beverly K. and Betzwieser, Joseph and Bhattacharjee, Dripta and Billingsley, Garilynn and Biscans, Sebastien and Blair, Carl D. and Blair, Ryan M. and Bode, Nina and Booker, Phillip and Bork, Rolf and Bramley, Alyssa and Brown, Daniel D. and Buikema, Aaron and Cahillane, Craig and Cannon, Kipp C. and Cao, Huy Tuong and Chen, Xu and Ciobanu, Alexei A. and Clara, Filiberto and Compton, Camilla and Cooper, Sam J. and Corley, Kenneth R. and Countryman, Stefan T. and Covas, Pep B. and Coyne, Dennis C. and Datrier, Laurence E. and Davis, Derek and Difronzo, Chiara D. and Dooley, Katherine L. and Driggers, Jenne C. and Dupej, Peter and Dwyer, Sheila E. and Effler, Anamaria and Etzel, Todd and Evans, Matthew and Evans, Tom M. and Feicht, Jon and Fernandez-Galiana, Alvaro and Fritschel, Peter and Frolov, Valery V. and Fulda, Paul and Fyffe, Michael and Giaime, Joe A. and Giardina, Dwayne D. and Godwin, Patrick and Goetz, Evan and Gras, Slawomir and Gray, Corey and Gray, Rachel and Green, Anna C. and Gupta, Anchal and Gustafson, Eric K. and Gustafson, Dick and Hall, Evan and Hanks, Jonathan and Hanson, Joe and Hardwick, Terra and Hasskew, Raine K. and Heintze, Matthew C. and Helmling-Cornell, Adrian F. and Holland, Nathan A. and Izmui, Kiamu and Jia, Wenxuan and Jones, Jeff D. and Kandhasamy, Shivaraj and Karki, Sudarshan and Kasprzack, Marie and Kawabe, Keita and Kijbunchoo, Nutsinee and King, Peter J. and Kissel, Jeffrey S. and Kumar, Rahul and Landry, Michael and Lane, Benjamin B. and Lantz, Brian and Laxen, Michael and Lecoeuche, Yannick K. and Leviton, Jessica and Jian, Liu and Lormand, Marc and Lundgren, Andrew P. and Macas, Ronaldas and Macinnis, Myron and Macleod, Duncan M. and Mansell, Georgia L. and Marka, Szabolcs and Marka, Zsuzsanna and Martynov, Denis V. and Mason, Ken and Massinger, Thomas J. and Matichard, Fabrice and Mavalvala, Nergis and McCarthy, Richard and McClelland, David E. and McCormick, Scott and McCuller, Lee and McIver, Jessica and McRae, Terry and Mendell, Gregory and Merfeld, Kara and Merilh, Edmond L. and Meylahn, Fabian and Mistry, Timesh and Mittleman, Richard and Moreno, Gerardo and Mow-Lowry, Conor M. and Mozzon, Simone and Mullavey, Adam and Nelson, Timothy J. and Nguyen, Philippe and Nuttall, Laura K. and Oberling, Jason and Oram, Richard J. and Osthelder, Charles and Ottaway, David J. and Overmier, Harry and Palamos, Jordan R. and Parker, William and Payne, Ethan and Pele, Arnaud and Penhorwood, Reilly and Perez, Carlos J. and Pirello, Marc and Radkins, Hugh and Ramirez, Karla E. and Richardson, Jonathan W. and Riles, Keith and Robertson, Norna A. and Rollins, Jameson G. and Romel, Chandra L. and Romie, Janeen H. and Ross, Michael P. and Ryan, Kyle and Sadecki, Travis and Sanchez, Eduardo J. and Sanchez, Luis E. and Tiruppatturrajamanikkam, Saravanan R. and Savage, Richard L. and Schaetzl, Dean and Schnabel, Roman and Schofield, Robert M. and Schwartz, Eyal and Sellers, Danny and Shaffer, Thomas and Sigg, Daniel and Slagmolen, Bram J. and Smith, Joshua R. and Soni, Siddharth and Sorazu, Borja and Spencer, Andrew P. and Strain, Ken A. and Sun, Ling and Szczepanczyk, Marek J. and Thomas, Michael and Thomas, Patrick and Thorne, Keith A. and Toland, Karl and Torrie, Calum I. and Traylor, Gary and Tse, Maggie and Urban, Alexander L. and Valdes, Guillermo and Vander-Hyde, Daniel C. and Veitch, Peter J. and Venkateswara, Krishna and Venugopalan, Gautam and Viets, Aaron D. and Vo, Thomas and Vorvick, Cheryl and Wade, Madeline and Ward, Robert L. and Warner, Jim and Weaver, Betsy and Weiss, Rainer and Whittle, Chris and Willke, Benno and Wipf, Christopher C. and Xiao, Liting and Yu, Hang and Yu, Haocun and Zhang, Liyuan and Zucker, Michael E. and Zweizig, John},
   year={2021},
   month=apr, pages={4047} }

@software{brown_2025_12662017,
  author       = {Brown, Daniel David and
                  Freise, Andreas and
                  Cao, Huy Tuong and
                  Ciobanu, Alexei and
                  Gobeil, Jeremie and
                  Green, Anna and
                  Hapke, Paul and
                  Jones, Philip and
                  van der Kolk, Miron and
                  Kuns, Kevin and
                  Leavey, Sean and
                  Perry, Jonathan Warren and
                  Rowlinson, Samuel and
                  Sallé, Mischa},
  title        = {FINESSE},
  month        = mar,
  year         = 2025,
  publisher    = {Gitlab},
  version      = {3.0a32},
  doi          = {10.5281/zenodo.12662017},
  url          = {https://doi.org/10.5281/zenodo.12662017},
}

@techreport{ETDesignReportUpdate2020,
  title        = {{ET Design Report Update 2020}},
  author       = {{ET Steering Committee}},
  institution  = {Einstein Telescope},
  type         = {Official document},
  number       = {ET-0007C-20},
  year         = {2024},
  month        = feb,
  note         = {Latest release. Code issue time: 13:22, 19 February 2024. Series: Projects ILIAS and Design Study Project, WP5 -- Management. Previous releases: ET-0007A-20, ET-0007B-20},
  url          = {https://apps.et-gw.eu/tds/ql/?c=15418}
}

@article{PhysRevD.90.122011,
  title = {Fabry-P\'erot-Michelson interferometer using higher-order Laguerre-Gauss modes},
  author = {Gatto, A. and Tacca, M. and K\'ef\'elian, F. and Buy, C. and Barsuglia, M.},
  journal = {Phys. Rev. D},
  volume = {90},
  issue = {12},
  pages = {122011},
  numpages = {9},
  year = {2014},
  month = {Dec},
  publisher = {American Physical Society},
  doi = {10.1103/PhysRevD.90.122011},
  url = {https://link.aps.org/doi/10.1103/PhysRevD.90.122011}
}

@article{Wang_2017,
   title={Thermal modelling of Advanced LIGO test masses},
   volume={34},
   ISSN={1361-6382},
   url={http://dx.doi.org/10.1088/1361-6382/aa6e60},
   DOI={10.1088/1361-6382/aa6e60},
   number={11},
   journal={Classical and Quantum Gravity},
   publisher={IOP Publishing},
   author={Wang, H and Blair, C and Dovale Álvarez, M and Brooks, A and Kasprzack, M F and Ramette, J and Meyers, P M and Kaufer, S and O’Reilly, B and Mow-Lowry, C M and Freise, A},
   year={2017},
   month=may, pages={115001} }

@techreport{AsharpTCS:2025,
    Author = {Huy-Tuong Cao and Kevin Kuns},
    Title = {{A$\sharp$ TCS Considerations for Heavy SUS (Review and Update)}},
    Type = {LIGO Technical Report / Presentation},
    Number = {LIGO-G2502579-v2},
    Year = {2025},
    Month = dec,
    Note = {Presentation at BHQS "Heavy SUS" Workshop \#4.},
    url = {https://dcc.ligo.org/LIGO-G2502579/}
}

@article{Rosauer:25,
author = {Tyler Rosauer and Huy Tuong Cao and Mohak Bhattacharya and Peter Carney and Luke Johnson and Shane Levin and Cynthia Liang and Xuesi Ma and Luis Martin Gutierrez and Michael Padilla and Liu Tao and Aiden Wilkin and Aidan Brooks and Jonathan W. Richardson},
journal = {Optica},
number = {10},
pages = {1569--1577},
publisher = {Optica Publishing Group},
title = {Demonstration of a next-generation wavefront actuator for gravitational-wave detection},
volume = {12},
month = {Oct},
year = {2025},
url = {https://opg.optica.org/optica/abstract.cfm?URI=optica-12-10-1569},
doi = {10.1364/OPTICA.567608},
}

@article{h91l-w8w7,
  title = {Transverse distance estimation with higher-order Hermite-Gauss modes},
  author = {Paneru, Dilip and D'Errico, Alessio and Karimi, Ebrahim},
  journal = {Phys. Rev. A},
  volume = {113},
  issue = {1},
  pages = {L011702},
  numpages = {7},
  year = {2026},
  month = {Jan},
  publisher = {American Physical Society},
  doi = {10.1103/h91l-w8w7},
  url = {https://link.aps.org/doi/10.1103/h91l-w8w7}
}

@ARTICLE{HOM_nature,
       author = {{Zou}, Kaiheng and {Pang}, Kai and {Song}, Hao and {Fan}, Jintao and {Zhao}, Zhe and {Song}, Haoqian and {Zhang}, Runzhou and {Zhou}, Huibin and {Minoofar}, Amir and {Liu}, Cong and {Su}, Xinzhou and {Hu}, Nanzhe and {McClung}, Andrew and {Torfeh}, Mahsa and {Arbabi}, Amir and {Tur}, Moshe and {Willner}, Alan E.},
        title = "{High-capacity free-space optical communications using wavelength- and mode-division-multiplexing in the mid-infrared region}",
      journal = {Nature Communications},
         year = 2022,
        month = dec,
       volume = {13},
          eid = {7662},
        pages = {7662},
          doi = {10.1038/s41467-022-35327-w},
       adsurl = {https://ui.adsabs.harvard.edu/abs/2022NatCo..13.7662Z}
}

@article{PhysRevLett.105.231102,
  title = {Higher-Order Laguerre-Gauss Mode Generation and Interferometry for Gravitational Wave Detectors},
  author = {Granata, M. and Buy, C. and Ward, R. and Barsuglia, M.},
  journal = {Phys. Rev. Lett.},
  volume = {105},
  issue = {23},
  pages = {231102},
  numpages = {4},
  year = {2010},
  month = {Nov},
  publisher = {American Physical Society},
  doi = {10.1103/PhysRevLett.105.231102},
  url = {https://link.aps.org/doi/10.1103/PhysRevLett.105.231102}
}

@article{PhysRevD.92.102002,
  title = {Higher-order Laguerre-Gauss interferometry for gravitational-wave detectors with in situ mirror defects compensation},
  author = {Allocca, A. and Gatto, A. and Tacca, M. and Day, R. A. and Barsuglia, M. and Pillant, G. and Buy, C. and Vajente, G.},
  journal = {Phys. Rev. D},
  volume = {92},
  issue = {10},
  pages = {102002},
  numpages = {9},
  year = {2015},
  month = {Nov},
  publisher = {American Physical Society},
  doi = {10.1103/PhysRevD.92.102002},
  url = {https://link.aps.org/doi/10.1103/PhysRevD.92.102002}
}

@article{PhysRevLett.93.250602,
  title = {Thermal-Noise Limit in the Frequency Stabilization of Lasers with Rigid Cavities},
  author = {Numata, Kenji and Kemery, Amy and Camp, Jordan},
  journal = {Phys. Rev. Lett.},
  volume = {93},
  issue = {25},
  pages = {250602},
  numpages = {4},
  year = {2004},
  month = {Dec},
  publisher = {American Physical Society},
  doi = {10.1103/PhysRevLett.93.250602},
  url = {https://link.aps.org/doi/10.1103/PhysRevLett.93.250602}
}

@book{DovaleAlvarez:2019ugw,
    author = "Dovale Alvarez, Miguel",
    title = "{Optical Cavities for Optical Atomic Clocks, Atom Interferometry and Gravitational-Wave Detection}",
    doi = "10.1007/978-3-030-20863-9",
    isbn = "978-3-030-20862-2, 978-3-030-20863-9",
    publisher = "Springer",
    series = "Springer Theses",
    year = "2019"
}

@article{Oelker:2019kqe,
    author = "Oelker, E. and others",
    title = "{Demonstration of 4.8{\,}{\texttimes}{\,}10{\ensuremath{-}}17 stability at 1{\,}s for two independent optical clocks}",
    eprint = "1902.02741",
    archivePrefix = "arXiv",
    primaryClass = "physics.atom-ph",
    doi = "10.1038/s41566-019-0493-4",
    journal = "Nature Photon.",
    volume = "13",
    number = "10",
    pages = "714--719",
    year = "2019"
}

@article{Savalle_2021,
   title={Searching for Dark Matter with an Optical Cavity and an Unequal-Delay Interferometer},
   volume={126},
   ISSN={1079-7114},
   url={http://dx.doi.org/10.1103/PhysRevLett.126.051301},
   DOI={10.1103/physrevlett.126.051301},
   number={5},
   journal={Physical Review Letters},
   publisher={American Physical Society (APS)},
   author={Savalle, Etienne and Hees, Aurélien and Frank, Florian and Cantin, Etienne and Pottie, Paul-Eric and Roberts, Benjamin M. and Cros, Lucie and McAllister, Ben T. and Wolf, Peter},
   year={2021},
   month=feb }

@misc{LG06_2026,
  author = {Tao, Liu and others },
  title  = {Improving Beam Quality in Gravitational-Wave Interferometers Illuminated by Higher-Order Laguerre-Gaussian Modes},
  year   = {2026},
  note   = {Under preparation}
}

\end{document}